# Lightweight authenticated quantum key distribution protocols with key recycling

**Jun Gu [1], Tzonelih Hwang[*]**

*Department of Computer Science and Information Engineering, National Cheng Kung University, No. 1, University Rd., Tainan City, 70101, Taiwan, R.O.C.*

[1] isgujun@163.com

[*]**Responsible for correspondence:**
Tzonelih Hwang
Distinguished Professor
Department of Computer Science and Information Engineering,
National Cheng Kung University,
No. 1, University Rd.,
Tainan City, 70101, Taiwan, R.O.C.
Email: hwangtl@csie.ncku.edu.tw
TEL: +886-6-2757575 ext. 62524

# Abstract

Quantum key distribution (QKD) has been developed for decades and several different QKD protocols have been proposed. But two difficulties limit the implementation of most QKD protocols. First, the involved participants are required to have heavy quantum capabilities, such as quantum joint operation, quantum register, and so on. Second, a hypothetical authenticated classical channel is used in most of the existing QKD protocols and this assumed channel does not exist in reality. To solve both the above limitations at the same time, this study proposes three lightweight authenticated QKD protocols with key recycling and shows these proposed protocols are robust under the collective attack.

**Keywords:** Quantum key distribution; Lightweight quantum key distribution; Collective attack

## 1. Introduction

In 1984, Bennet et al. [1] proposed the first quantum key distribution (QKD) protocol to help the involved participants share an unconditionally secure key. Subsequently, several QKD protocols [2-8] have been proposed in decades. However, most of these QKD protocols [2-5] have two difficulties in implementation. First, these QKD protocols need all the participants to have heavy quantum capabilities, such as quantum joint operation, quantum register, and so on. That means a participant with only limited quantum capabilities cannot be involved in these protocols. Second, an assumed authenticated classical channel is the prerequisite for running these protocols. That is, all the above QKD protocols adopt an ideally authenticated classical channel where the transmitted information cannot be modified and the identities of the communicating parties cannot be impersonated.

To solve the first problem, Boyer et al. proposed two semi-quantum key distribution (SQKD) protocols [9] where a classical participant who just has restricted quantum capacities can be involved. In Boyer et al.'s SQKD protocols, the classical participants are restricted to perform three out of the following four operations: (1) preparing qubits in Z-basis $\{|0\rangle,|1\rangle\}$, (2) measuring qubits with the Z-basis, (3) reordering the qubits via different quantum delay lines, and (4) reflecting the qubits. Afterward, various types of semi-quantum protocols [10-14] have been proposed. However, Julsgaard et al.'s study [15] showed that the photons are difficult to be held, even for a short time. Hence, the operation of reordering appears to be quite difficult for implementation. To make the semi-quantum protocols easier for implementation, Hwang et al. [16] proposed the definition of lightweight quantum protocol. Lightweight quantum protocol allows the lightweight participants to be involved in the protocol. Here, the lightweight participants are restricted to perform only two out of the following four operations: (1) preparing qubits in Z-basis, (2) measuring qubits with the Z-basis, (3) performing single-photon unitary operations, and (4) reflecting the qubits. According to this definition, several well-known protocols can be considered as lightweight quantum protocols, such as BB84 [1], E91 [8], and several measurement-device-independent protocols [6, 7].

To solve the second problem, several authenticated quantum protocols have been proposed. Different from the above quantum protocols needing an assumed ideal authenticated classical channel, the authenticated quantum protocols use pre-shared keys to help the participants authenticate with each other. Besides, compared with the classical authentication protocols, the quantum authentication protocols has an advantage that the pre-shared keys are protected by quantum mechanics [17]. If there is no eavesdropper been detected, the pre-shared keys can be reused as newly pre-shared unconditional secure keys. However, if an eavesdropper is detected, most of the authenticated quantum protocols have to discard all the pre-shared keys and the participants should share new keys again for running the protocols. And that could be very complicate and also difficult in implementation. To solve it, an existing concept of quantum key recycling [18-20] can be used for designing the authenticated quantum protocols. That is, even an eavesdropper has been detected, parts of the pre-shared keys still can be reused as unconditional secure pre-shared keys.

This study proposes three lightweight authenticated quantum key distribution (LAQKD) protocols where both the above mentioned lightweight property and authentication property can be achieved simultaneously. With these proposed protocols, the participants are allowed to have various lightweight quantum capabilities. The proposed LAQKD protocols can be suitable for many scenarios in reality. That means, the participants do not need to change their quantum capabilities for adapting a special QKD protocol anymore. They can choose an appropriate LAQKD protocol according to their own quantum capabilities instead.

Besides, different from almost all the existing authenticated quantum protocols where the pre-shared master keys have to be discarded encountering an eavesdropping, most of the pre-shared keys used in the proposed LAQKD protocols can be recycled even when an eavesdropping is detected. For each proposed LAQKD protocol, a key recycling rate analysis is given to show the recycling threshold of each pre-shared key.

Moreover, to demonstrate the pros and cons of the proposed LAQKD protocols as compared to several existing QKD protocols more comprehensively, a concept of transmission time cost for quantum protocol is first introduced and defined in this study. That is, for most of the existing studies, the designing of a quantum protocol just focus on getting a better qubit efficiency, needing fewer quantum capabilities or using some quantum resources which are easier for implementation. However, to achieve these requirements, some extra costs have to be paid and these extra costs have never been shown in the comparison part. The missing part makes it difficult to truly show the pros and cons of each protocol. In this study, a new concept named transmission time cost is defined to make the comparison more comprehensive.

The rest of this paper is organized as follows. Section 2 shows the details of three proposed LAQKD protocols. Section 3 uses collective attack analysis to show that the proposed protocols are robust. Section 4 analyzes the expectation of the key recycling rate of the proposed protocols. Section 5 first introduces the definition of transmission time cost and then compares the proposed protocols with several well-known QKD protocols. At last, a conclusion is given in Section 6.

## 2. Three LAQKD protocols

In this section, three LAQKD protocols are proposed and described one by one.

### 2.1 Protocol 1: LAQKD protocol with generation and unitary operation

Before describing the proposed Protocol 1, we first introduce the environment adopted in it. There are two participants Alice and Bob, who just have two lightweight quantum capabilities: generating Z-basis qubits and performing single photon unitary operations, involved in the protocol. They try to use an $n+m$-bits pre-shared master key $K_1 = \{k_1^1, k_1^2, \cdots, k_1^{n+m}\}$ to share a secure session key $K$ of $n'$ bits with the help of an untrusted third party (TP). Here, $n$ is the length of the raw key further shared in the protocol, $m$ is the length of a hash function [21] output and $n'$ is the number of bits extracted from $n$ by performing a private amplification on the raw key. Besides, a backup master key $K_1' = \{k_1'^1, k_1'^2, \cdots, k_1'^l\}$ is also pre-shared between Alice and Bob where $l$ is a large enough number. The $K_1'$ is used for making up the length of $K_1$. That is, if an eavesdropper is detected during the execution of this protocol, parts of $K_1$ will be discarded and parts of it can be recycled for further use. However, the length of the recycling part may be insufficient. To solve this problem, parts of $K'_1$ can be used for making up the length of $K_1$.

In the proposed Protocol 1, both the quantum channels and the classical channels used are noiseless, but all the transmitted qubits and classical bits on them can be modified by anyone. Then, the proposed Protocol 1 can be described step by step as follows (Figure 1):

**Step 1**: Alice (Bob) generates a random bit sequence $R_A = \{r_A^1, r_A^2, \cdots, r_A^n\}$ ($R_B = \{r_B^1, r_B^2, \cdots, r_B^n\}$) and then performs a hash function [21] on $R_A$ ($R_B$) to obtain $h(R_A)$ ($h(R_B)$).

**Step 2**: Alice (Bob) generates $n+m$ Z-basis single photons to form an ordered qubit sequence $Q_A = \{q_A^1, q_A^2, \cdots, q_A^{n+m}\}$ ($Q_B = \{q_B^1, q_B^2, \cdots, q_B^{n+m}\}$) according to $R_A \| h(R_A)$ ($R_B \| h(R_B)$). That is, if the $i$th bit in $R_A \| h(R_A)$ ($R_B \| h(R_B)$) is '0', then the $i$th qubit in $Q_A$ ($Q_B$) is $|0\rangle$. Otherwise, the $i$th qubit is $|1\rangle$. Afterward, Alice (Bob) performs unitary operations $I = |0\rangle\langle 0| + |1\rangle\langle 1|$ or $H = \frac{1}{\sqrt{2}}(|0\rangle\langle 0| + |0\rangle\langle 1| + |1\rangle\langle 0| - |1\rangle\langle 1|)$ on $Q_A$ ($Q_B$) according to $K_1$ to obtain $Q'_A$ ($Q'_B$). That is, if $k_1^i = 0$, Alice and Bob perform the operation $I$ on $q_A^i$ and $q_B^i$, respectively. Otherwise, an operation $H$ is performed on $q_A^i$ and $q_B^i$. Finally, Alice and Bob send $Q'_A$ and $Q'_B$ to TP.

**Step 3**: After TP receives $Q'_A$ and $Q'_B$, he/she performs Bell measurement on each qubit pair $\{q'^i_A, q'^i_B\}$ in order and announces the measurement results.

**Step 4:** As shown in Table 1, Alice and Bob deduce a bit sequence

$M = \{m^1, m^2, \cdots, m^{n+m}\}$ from the measurement results. For example, if $k_1^i = 0$ and the measurement result is $|\Psi^-\rangle$, then $m^i = 1$. Here, we can find that $M$ is expected to be equal to $(R_A \oplus R_B) \| (h(R_A) \oplus h(R_B))$.

**Step 5:** Alice (Bob) computes the equation $M \oplus (R_A \| h(R_A))$ ($M \oplus (R_B \| h(R_B))$) to obtain $R_B' \| h(R_B)'$ ($R_A' \| h(R_A)'$). Then, Alice (Bob) performs the hash function on the obtained $R_B'$ ($R_A'$) and checks whether $h(R_B')$ ($h(R_A')$) is equal to $h(R_B)'$ ($h(R_A)'$). If they are not equal, there must be an eavesdropper during the qubit transmission processes. Then, this protocol will be aborted. Otherwise, Alice and Bob extract the session key $K$ by performing privacy amplification [22] on the raw key $R_A$.

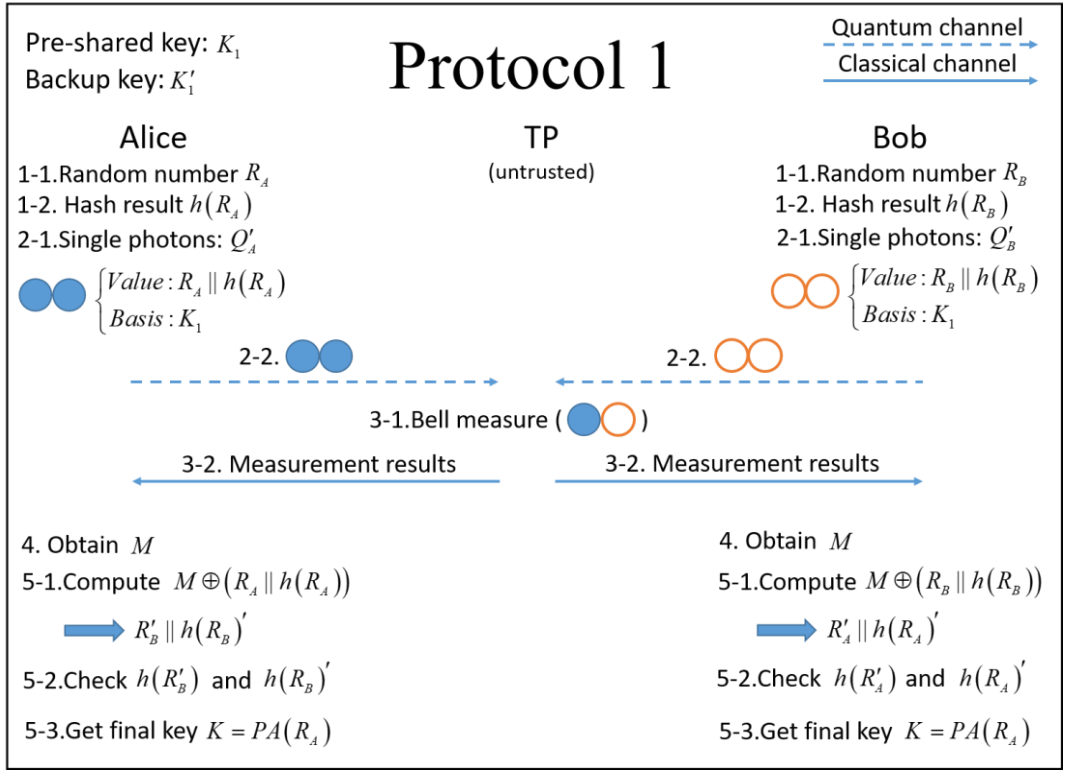

Figure 1. Protocol 1: LAQKD protocol with generation and unitary operation

Table 1. The relationship of each bit, qubit and measurement result in Protocol 1

| $K_1$ | $R_A \| h(R_A), R_B \| h(R_B)$ | $Q'_A, Q'_B$ | Measurement result | $M$ |
|---|---|---|---|---|
| | 0 , 0 | $\|0\rangle, \|0\rangle$ | $\|\Phi^+\rangle$ or $\|\Phi^-\rangle$ | 0 |
| | 0 , 1 | $\|0\rangle, \|1\rangle$ | $\|\Psi^+\rangle$ or $\|\Psi^-\rangle$ | 1 |
| 0 | 1 , 0 | $\|1\rangle, \|0\rangle$ | $\|\Psi^+\rangle$ or $\|\Psi^-\rangle$ | 1 |
| | 1 , 1 | $\|1\rangle, \|1\rangle$ | $\|\Phi^+\rangle$ or $\|\Phi^-\rangle$ | 0 |
| | 0 , 0 | $\|+\rangle, \|+\rangle$ | $\|\Phi^+\rangle$ or $\|\Psi^+\rangle$ | 0 |
| | 0 , 1 | $\|+\rangle, \|-\rangle$ | $\|\Phi^-\rangle$ or $\|\Psi^-\rangle$ | 1 |
| 1 | 1 , 0 | $\|-\rangle, \|+\rangle$ | $\|\Phi^-\rangle$ or $\|\Psi^-\rangle$ | 1 |
| | 1 , 1 | $\|-\rangle, \|-\rangle$ | $\|\Phi^+\rangle$ or $\|\Psi^+\rangle$ | 0 |

In this proposed protocol, because the participants can check whether the announced Bell measurement results are correct or not by themselves according to Table 1. Hence, any eavesdropper, even the TP, can be detected if some malicious behavior occurs. As a result, it implies that TP can be considered to be untrusted. A formal proof of robustness under collective attack will be given in Section 3.

**2.2 Protocol 2: LAQKD with measurement and unitary operation**

Different from Protocol 1, in Protocol 2, Alice and Bob just have the lightweight quantum capabilities of measuring qubits with Z-basis and performing single photon unitary operations. Moreover, two master keys $K_1 = \{k_1^1, k_1^2, \cdots, k_1^n\}$ and $K_2 = \{k_2^1, k_2^2, \cdots, k_2^n\}$ are pre-shared between Alice and Bob. Two backup master keys $K'_1 = \{k_1'^1, k_1'^2, \cdots, k_1'^n\}$ and $K'_2 = \{k_2'^1, k_2'^2, \cdots, k_2'^n\}$ are also pre-shared.

The same as Protocol 1, an untrusted TP is involved in the protocol and both the quantum channels and the classical channels used in this protocol are noiseless. Then, the proposed Protocol 2 can be described as follows (Figure 2):

**Step 1'**: TP generates $n$ Bell states $B=\{(q_A^1,q_B^1),(q_A^2,q_B^2),\cdots,(q_A^n,q_B^n)\}$ in $|\Phi^-\rangle=\frac{1}{\sqrt{2}}(|00\rangle-|11\rangle)=\frac{1}{\sqrt{2}}(|+-\rangle+|-+\rangle)$ and picks out all the first particles and all the second particles in $B$ to form two ordered particle sequences $Q_A=\{q_A^1,q_A^2,\cdots,q_A^n\}$ and $Q_B=\{q_B^1,q_B^2,\cdots,q_B^n\}$, respectively. Then, he/she sends $Q_A$ and $Q_B$ to Alice and Bob, respectively.

**Step 2'**: Alice (Bob) performs unitary operations $I$ or $H$ on $Q_A$ ($Q_B$) according to $K_1$ to obtain $Q'_A$ ($Q'_B$). Here, the performing rules are the same as in Step 2 of the first proposed protocol. Then, Alice (Bob) uses Z-basis to measure each qubit in $Q'_A$ ($Q'_B$) to obtain $R_A=\{r_A^1,r_A^2,\cdots,r_A^n\}$ ($R_B=\{r_B^1,r_B^2,\cdots,r_B^n\}$). Finally, according to Table 2, Alice deduces $R=\{r^1,r^2,\cdots,r^n\}=R_A$ and Bob deduces $R=R_B\oplus K_1$.

**Step 3'**: Alice and Bob divide $R$ into two bit sequences $R_0$ and $R_1$ according to $K_2$. That is, if $k_2^i=0$, then $r^i$ belongs to $R_0$. Otherwise, $r^i$ belongs to $R_1$. Subsequently, Alice (Bob) performs a hash function on $R_0$ ($R_1$) to obtain $h(R_0)$ ($h(R_1)$). Then, Alice (Bob) sends $h(R_0)$ ($h(R_1)$) to Bob (Alice).

**Step 4'**: After receiving $h(R_1)$ ($h(R_0)$), Alice (Bob) performs the hash function on the held $R_1$ ($R_0$) to check whether the received hash function result is equal to the computational one. If they are not equal, there must be an eavesdropper during the qubit transmission processes. Then, this protocol will be aborted. Otherwise, Alice and Bob extract the session key $K$ by performing privacy amplification on $R$.

Table 2. The relationship of each bit, qubit and measurement result in Protocol 2

| $\{Q_A, Q_B\}$ | $K_1$ | Measurement result | $R_A$ | $R_B$ | $R$ |
|---|---|---|---|---|---|
| $\|\Phi^-\rangle$ | 0 | $\|0\rangle_{Q'_A}\|0\rangle_{Q'_B}$ | 0 | 0 | 0 |
| | | $\|1\rangle_{Q'_A}\|1\rangle_{Q'_B}$ | 1 | 1 | 1 |
| | 1 | $\|0\rangle_{Q'_A}\|1\rangle_{Q'_B}$ | 0 | 1 | 0 |
| | | $\|1\rangle_{Q'_A}\|0\rangle_{Q'_B}$ | 1 | 0 | 1 |

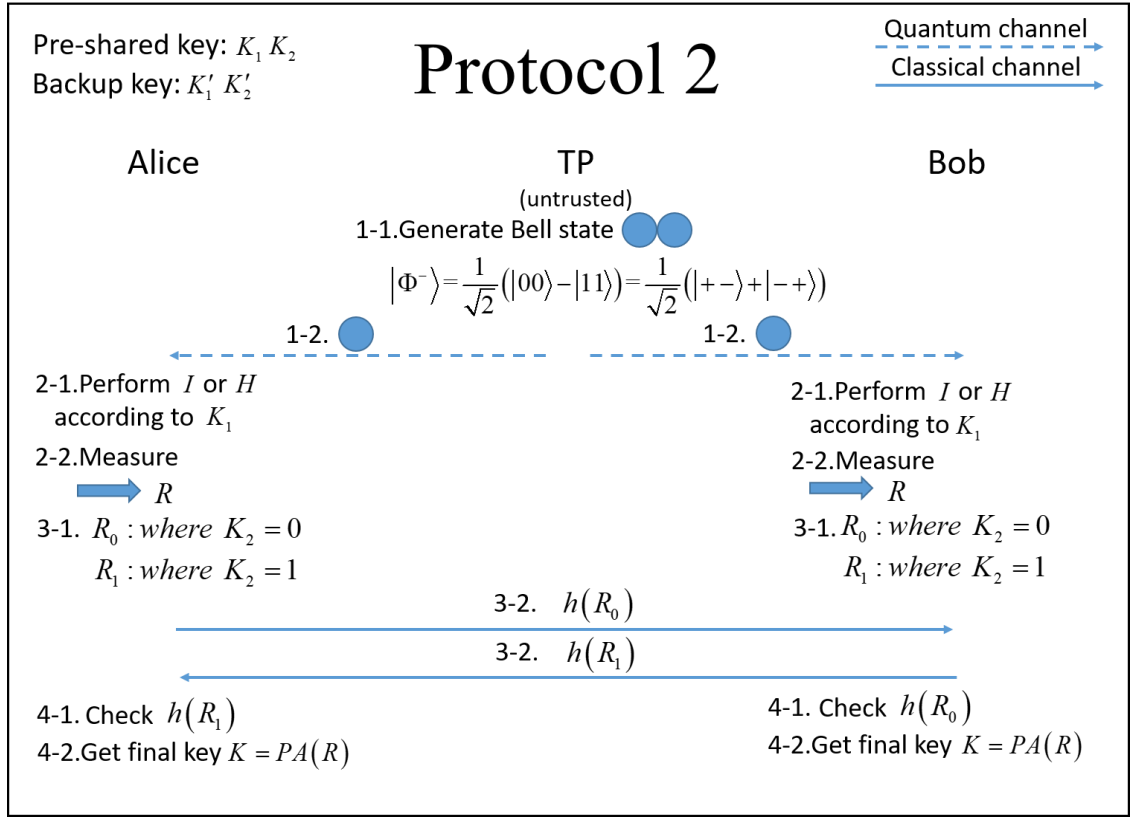

Figure 2. Protocol 2: LAQKD with measurement and unitary operation

In the proposed LAQKD protocol 2, the eavesdropping detection is done by detecting whether the initial state is indeed $|\Phi^-\rangle$ or not. Moreover, the participants can check the initial states by themselves. Therefore, any eavesdropper, even the TP, can be detected if some malicious behavior occurs. As a result, it means that TP can be considered to be untrusted. A formal proof of robustness under collective attack will be given in Section 3.

**2.3 Protocol 3: LAQKD with reflection and unitary operation**

In Protocol 3, Alice and Bob just have the lightweight quantum capabilities of reflecting qubits and performing single photon unitary operations. The master key $K_1=\{k_1^1,k_1^2,\cdots,k_1^{n+m}\}$ and the backup master key $K_1'=\{k_1'^1,k_1'^2,\cdots,k_1'^l\}$ are pre-shared between Alice and Bob. The same as the above two protocols, the quantum channels and the classical channels used in this protocol are noiseless and an untrusted TP is involved in the protocol. Moreover, a unitary operation $H'=\frac{1}{\sqrt{2}}(|0\rangle\langle 0|-|0\rangle\langle 1|+|1\rangle\langle 0|+|1\rangle\langle 1|)$ is used in this protocol. The $H'$ can transform the qubits $\{|0\rangle,|1\rangle,|+\rangle,|-\rangle\}$ from one to the other. The transformation relationship is shown as follows:

$$\begin{cases} |0\rangle \xrightarrow{H'} |+\rangle \\ |1\rangle \xrightarrow{H'} |-\rangle \\ |+\rangle \xrightarrow{H'} |1\rangle \\ |-\rangle \xrightarrow{H'} |0\rangle \end{cases}$$

The proposed Protocol 3 can be described as follows (Figure 3):

**Step 1\***: TP generates $2(n+m)$ single photons in the state $|0\rangle$ and divides them into two ordered qubit sequences $Q_A=\{q_A^1,q_A^2,\cdots,q_A^{n+m}\}$ and $Q_B=\{q_B^1,q_B^2,\cdots,q_B^{n+m}\}$. Then, he/she sends $Q_A$ and $Q_B$ to Alice and Bob, respectively.

**Step 2\***: Alice (Bob) generates an $n$-bits random number $R_A=\{r_A^1,r_A^2,\cdots,r_A^n\}$

($R_B = \{r_B^1, r_B^2, \cdots, r_B^n\}$) and performs a hash function on $R_A$ ($R_B$) to obtain $h(R_A)$ ($h(R_B)$).

**Step 3\*:** Alice (Bob) performs $H'^j$, the $j\,(j \in \{0,1,2,3\})$ times unitary operation $H'$ on $Q_A$ ($Q_B$) according to $R_A \| h(R_A)$ ($R_B \| h(R_B)$) and $K_1$ (as shown in Table 3). Here, for the $i$-th qubit in $Q_A$ ($Q_B$), $j = k_1^i + 2r_A^i$ ($j = k_1^i + 2r_B^i$). For example, if $k_1^i = 1$ and $r_A^i = 1$, then Alice performs 3 times $H'$ on $Q_A$. After this, the qubit sequences $Q_A$ ($Q_B$) are transformed into $Q_A'$ ($Q_B'$). Then, Alice (Bob) sends $Q_A'$ ($Q_B'$) back to TP.

**Step 4\*:** Upon receiving $Q_A'$ and $Q_B'$, TP performs Bell measurement on each qubit pair $\{q_A'^i, q_B'^i\}$ and announces the measurement results.

**Step 5\*:** According to Table 3, Alice and Bob can obtain a bit sequence $M = \{m^1, m^2, \cdots, m^{n+m}\}$ from the measurement results. For example, if $k_1^i = 0$ and the measurement result is $|\Psi^-\rangle$, then $m^i = 1$. Here, we can find that $M$ is expected to be equal to $(R_A \oplus R_B) \| (h(R_A) \oplus h(R_B))$.

**Step 6\*:** (The same as Step 5) Alice (Bob) computes the equation $M \oplus (R_A \| h(R_A))$ ($M \oplus (R_B \| h(R_B))$) to obtain $R_B' \| h(R_B)'$ ($R_A' \| h(R_A)'$). Then, Alice (Bob) performs the hash function on the obtained $R_B'$ ($R_A'$) and checks whether $h(R_B')$ ($h(R_A')$) is equal to $h(R_B)'$ ($h(R_A)'$). If they are not equal, there must be an eavesdropper during the qubit transmission processes. Then, this protocol will be aborted. Otherwise, Alice and Bob extract the session key $K$ by performing privacy amplification on the raw key $R_A$.

Table 3. The relationship of each bit, qubit and measurement result in Protocol 3

| $Q_A Q_B$ | $K_1$ | $R_A, R_B$ | $H'_A, H'_B$ | $Q'_A Q'_B$ | Measurement result | $M$ |
|---|---|---|---|---|---|---|
| $\|0\rangle_{Q_A}\|0\rangle_{Q_B}$ | 0 | 0 , 0 | $H'^0, H'^0$ | $\|0\rangle_{Q_A}\|0\rangle_{Q_B}$ | $\|\Phi^+\rangle$ or $\|\Phi^-\rangle$ | 0 |
| | | 0 , 1 | $H'^0, H'^2$ | $\|0\rangle_{Q_A}\|1\rangle_{Q_B}$ | $\|\Psi^+\rangle$ or $\|\Psi^-\rangle$ | 1 |
| | | 1 , 0 | $H'^2, H'^0$ | $\|1\rangle_{Q_A}\|0\rangle_{Q_B}$ | $\|\Psi^+\rangle$ or $\|\Psi^-\rangle$ | 1 |
| | | 1 , 1 | $H'^2, H'^2$ | $\|1\rangle_{Q_A}\|1\rangle_{Q_B}$ | $\|\Phi^+\rangle$ or $\|\Phi^-\rangle$ | 0 |
| | 1 | 0 , 0 | $H'^1, H'^1$ | $\|+\rangle_{Q_A}\|+\rangle_{Q_B}$ | $\|\Phi^+\rangle$ or $\|\Psi^+\rangle$ | 0 |
| | | 0 , 1 | $H'^1, H'^3$ | $\|+\rangle_{Q_A}\|-\rangle_{Q_B}$ | $\|\Phi^-\rangle$ or $\|\Psi^-\rangle$ | 1 |
| | | 1 , 0 | $H'^3, H'^1$ | $\|-\rangle_{Q_A}\|+\rangle_{Q_B}$ | $\|\Phi^-\rangle$ or $\|\Psi^-\rangle$ | 1 |
| | | 1 , 1 | $H'^3, H'^3$ | $\|-\rangle_{Q_A}\|-\rangle_{Q_B}$ | $\|\Phi^+\rangle$ or $\|\Psi^+\rangle$ | 0 |

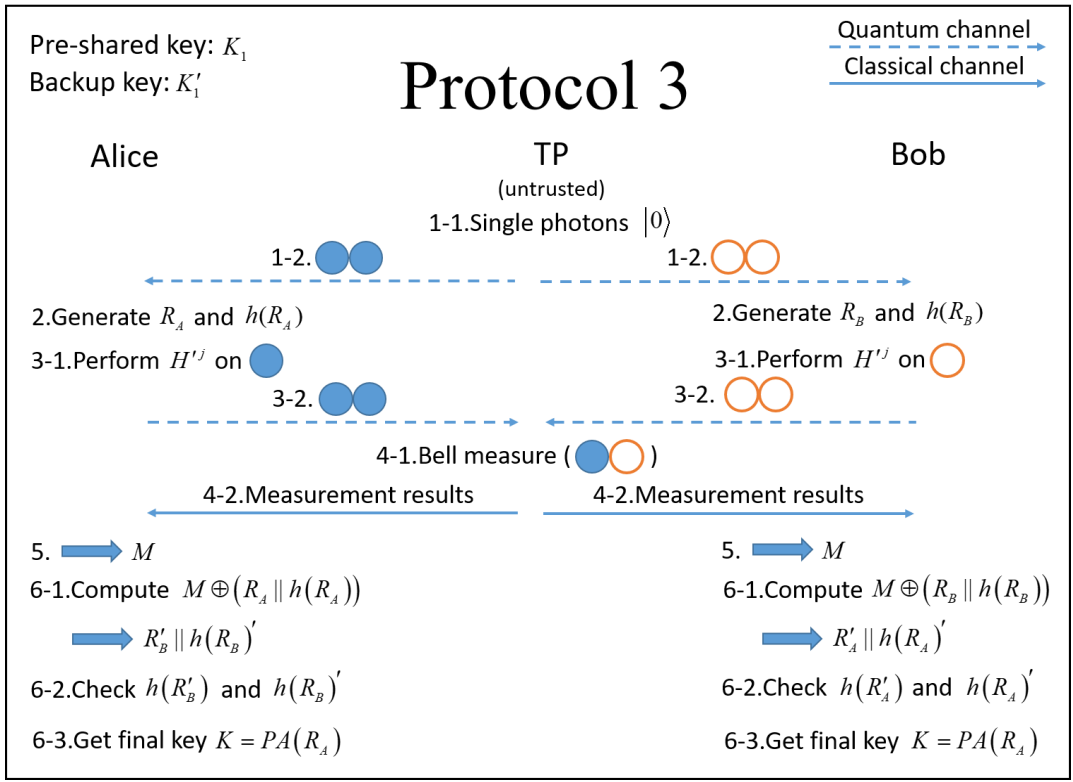

Figure 3. Protocol 3: LAQKD with reflection and unitary operation

In this protocol, any malicious behavior, even from the TP, can be detected by the participants by checking the announced Bell measurement results. Hence, the TP can be considered to be untrusted here. A formal proof of robustness under collective attack will be given in Section 3.

## 3. Security analysis

This section proves that the three proposed LAQKD protocols are robust. For the proposed LAQKD protocols, the robustness [9] means that any eavesdropper attempting to obtain the final shared key will be detected by the participants. To prove the robustness of the proposed LAQKD protocols, the collective attack analysis of each proposed LAQKD protocol is given. The collective attack is the strongest type of joint attack which also can be considered as the most general attack [23]. In the three proposed LAQKD protocols, TP can be considered as the strongest eavesdropper. Hence, if TP cannot obtain the final shared key or the pre-shared keys, then no other eavesdropper can do that. In the collective attack [23], an eavesdropper can be defined as follows.

(1) An eavesdropper can entangle her/his probe quantum systems with participants' quantum systems which are transmitted on the quantum channels and then tries to use the measurement results of his/her own probe quantum systems to obtain the final shared key or the pre-shared keys.

(2) Each quantum system may be transmitted several times in a protocol. For each time, an eavesdropper can perform a joint unitary operation on both the transmitted quantum system and his/her probe quantum system.

(3) An eavesdropper can keep his/her probe quantum systems until any later point in time. That means, he/she can choose to measure his/her probe quantum systems after obtaining some information coming from the announced information in the protocol.

According to these definitions, we show that all the proposed LAQKD protocols are robust by the following theorems and proofs.

**Theorem 1**: In the proposed LAQKD protocol 1, no eavesdropper can obtain the final shared key $K$ or the pre-shared key $K_1$ by the collective attack without being detected.

**Proof**: Assume TP is an eavesdropper. After Alice and Bob send their quantum systems $\{Q'_A, Q'_B\}$ to TP in Step 2, TP generates his/her probe quantum systems $E=\{e^1=|E_1\rangle, e^2=|E_2\rangle, \cdots, e^{n+m}=|E_{n+m}\rangle\}$ and performs a joint unitary operation $U_E$ on each pair of $\{q'^i_A, q'^i_B, e^i\}$ to entangle them. According to Table 1, we can find that $\{q'^i_A, q'^i_B\}$ has eight cases as follows.

Case 1: $\{q'^i_A, q'^i_B\}=\{|0\rangle_A, |0\rangle_B\}$; Case 2: $\{q'^i_A, q'^i_B\}=\{|0\rangle_A, |1\rangle_B\}$;

Case 3: $\{q'^i_A, q'^i_B\}=\{|1\rangle_A, |0\rangle_B\}$; Case 4: $\{q'^i_A, q'^i_B\}=\{|1\rangle_A, |1\rangle_B\}$;

Case 5: $\{q'^i_A, q'^i_B\}=\{|+\rangle_A, |+\rangle_B\}$; Case 6: $\{q'^i_A, q'^i_B\}=\{|+\rangle_A, |-\rangle_B\}$;

Case 7: $\{q'^i_A, q'^i_B\}=\{|-\rangle_A, |+\rangle_B\}$; Case 8: $\{q'^i_A, q'^i_B\}=\{|-\rangle_A, |-\rangle_B\}$;

Based on the eight cases, $U_E(q'^i_A, q'^i_B, e^i)$ can be described as follows:

Case 1:
$$U_E(|0\rangle_A \otimes |0\rangle_B \otimes |E_i\rangle_E) \\ = \alpha^{00}_{00}|\Phi^+\rangle_{AB}|e^{00}_{00}\rangle_E + \alpha^{01}_{00}|\Phi^-\rangle_{AB}|e^{01}_{00}\rangle_E + \alpha^{10}_{00}|\Psi^+\rangle_{AB}|e^{10}_{00}\rangle_E + \alpha^{11}_{00}|\Psi^-\rangle_{AB}|e^{11}_{00}\rangle_E \tag{1}$$

Case 2:
$$U_E(|0\rangle_A \otimes |1\rangle_B \otimes |E_i\rangle_E) \\ = \alpha^{00}_{01}|\Phi^+\rangle_{AB}|e^{00}_{01}\rangle_E + \alpha^{01}_{01}|\Phi^-\rangle_{AB}|e^{01}_{01}\rangle_E + \alpha^{10}_{01}|\Psi^+\rangle_{AB}|e^{10}_{01}\rangle_E + \alpha^{11}_{01}|\Psi^-\rangle_{AB}|e^{11}_{01}\rangle_E \tag{2}$$

Case 3:

$$U_E\left(|1\rangle_A \otimes |0\rangle_B \otimes |E_i\rangle_E\right)$$
$$= \alpha_{10}^{00}|\Phi^+\rangle_{AB}|e_{10}^{00}\rangle_E + \alpha_{10}^{01}|\Phi^-\rangle_{AB}|e_{10}^{01}\rangle_E + \alpha_{10}^{10}|\Psi^+\rangle_{AB}|e_{10}^{10}\rangle_E + \alpha_{10}^{11}|\Psi^-\rangle_{AB}|e_{10}^{11}\rangle_E \quad (3)$$

Case 4:

$$U_E\left(|1\rangle_A \otimes |1\rangle_B \otimes |E_i\rangle_E\right)$$
$$= \alpha_{11}^{00}|\Phi^+\rangle_{AB}|e_{11}^{00}\rangle_E + \alpha_{11}^{01}|\Phi^-\rangle_{AB}|e_{11}^{01}\rangle_E + \alpha_{11}^{10}|\Psi^+\rangle_{AB}|e_{11}^{10}\rangle_E + \alpha_{11}^{11}|\Psi^-\rangle_{AB}|e_{11}^{11}\rangle_E \quad (4)$$

Case 5:

$$U_E\left(|+\rangle_A \otimes |+\rangle_B \otimes |E_i\rangle_E\right)$$
$$= \frac{1}{2}U_E\left(|00E_i\rangle_{ABE} + |01E_i\rangle_{ABE} + |10E_i\rangle_{ABE} + |11E_i\rangle_{ABE}\right)$$

$$= \frac{1}{2}\begin{pmatrix} \alpha_{00}^{00}|\Phi^+\rangle_{AB}|e_{00}^{00}\rangle_E + \alpha_{00}^{01}|\Phi^-\rangle_{AB}|e_{00}^{01}\rangle_E + \alpha_{00}^{10}|\Psi^+\rangle_{AB}|e_{00}^{10}\rangle_E + \alpha_{00}^{11}|\Psi^-\rangle_{AB}|e_{00}^{11}\rangle_E \\ +\alpha_{01}^{00}|\Phi^+\rangle_{AB}|e_{01}^{00}\rangle_E + \alpha_{01}^{01}|\Phi^-\rangle_{AB}|e_{01}^{01}\rangle_E + \alpha_{01}^{10}|\Psi^+\rangle_{AB}|e_{01}^{10}\rangle_E + \alpha_{01}^{11}|\Psi^-\rangle_{AB}|e_{01}^{11}\rangle_E \\ +\alpha_{10}^{00}|\Phi^+\rangle_{AB}|e_{10}^{00}\rangle_E + \alpha_{10}^{01}|\Phi^-\rangle_{AB}|e_{10}^{01}\rangle_E + \alpha_{10}^{10}|\Psi^+\rangle_{AB}|e_{10}^{10}\rangle_E + \alpha_{10}^{11}|\Psi^-\rangle_{AB}|e_{10}^{11}\rangle_E \\ +\alpha_{11}^{00}|\Phi^+\rangle_{AB}|e_{11}^{00}\rangle_E + \alpha_{11}^{01}|\Phi^-\rangle_{AB}|e_{11}^{01}\rangle_E + \alpha_{11}^{10}|\Psi^+\rangle_{AB}|e_{11}^{10}\rangle_E + \alpha_{11}^{11}|\Psi^-\rangle_{AB}|e_{11}^{11}\rangle_E \end{pmatrix}$$

$$= \frac{1}{2}\begin{bmatrix} |\Phi^+\rangle_{AB}\left(\alpha_{00}^{00}|e_{00}^{00}\rangle_E + \alpha_{01}^{00}|e_{01}^{00}\rangle_E + \alpha_{10}^{00}|e_{10}^{00}\rangle_E + \alpha_{11}^{00}|e_{11}^{00}\rangle_E\right) \\ +|\Phi^-\rangle_{AB}\left(\alpha_{00}^{01}|e_{00}^{01}\rangle_E + \alpha_{01}^{01}|e_{01}^{01}\rangle_E + \alpha_{10}^{01}|e_{10}^{01}\rangle_E + \alpha_{11}^{01}|e_{11}^{01}\rangle_E\right) \\ +|\Psi^+\rangle_{AB}\left(\alpha_{00}^{10}|e_{00}^{10}\rangle_E + \alpha_{01}^{10}|e_{01}^{10}\rangle_E + \alpha_{10}^{10}|e_{10}^{10}\rangle_E + \alpha_{11}^{10}|e_{11}^{10}\rangle_E\right) \\ +|\Psi^-\rangle_{AB}\left(\alpha_{00}^{11}|e_{00}^{11}\rangle_E + \alpha_{01}^{11}|e_{01}^{11}\rangle_E + \alpha_{10}^{11}|e_{10}^{11}\rangle_E + \alpha_{11}^{11}|e_{11}^{11}\rangle_E\right) \end{bmatrix} \quad (5)$$

Case 6:

$$U_E\left(|+\rangle_A \otimes |-\rangle_B \otimes |E_i\rangle_E\right)$$
$$= \frac{1}{2}U_E\left(|00E\rangle_{ABE} - |01E\rangle_{ABE} + |10E\rangle_{ABE} - |11E\rangle_{ABE}\right)$$

$$= \frac{1}{2}\begin{pmatrix} \alpha_{00}^{00}|\Phi^+\rangle_{AB}|e_{00}^{00}\rangle_E + \alpha_{00}^{01}|\Phi^-\rangle_{AB}|e_{00}^{01}\rangle_E + \alpha_{00}^{10}|\Psi^+\rangle_{AB}|e_{00}^{10}\rangle_E + \alpha_{00}^{11}|\Psi^-\rangle_{AB}|e_{00}^{11}\rangle_E \\ -\alpha_{01}^{00}|\Phi^+\rangle_{AB}|e_{01}^{00}\rangle_E - \alpha_{01}^{01}|\Phi^-\rangle_{AB}|e_{01}^{01}\rangle_E - \alpha_{01}^{10}|\Psi^+\rangle_{AB}|e_{01}^{10}\rangle_E - \alpha_{01}^{11}|\Psi^-\rangle_{AB}|e_{01}^{11}\rangle_E \\ +\alpha_{10}^{00}|\Phi^+\rangle_{AB}|e_{10}^{00}\rangle_E + \alpha_{10}^{01}|\Phi^-\rangle_{AB}|e_{10}^{01}\rangle_E + \alpha_{10}^{10}|\Psi^+\rangle_{AB}|e_{10}^{10}\rangle_E + \alpha_{10}^{11}|\Psi^-\rangle_{AB}|e_{10}^{11}\rangle_E \\ -\alpha_{11}^{00}|\Phi^+\rangle_{AB}|e_{11}^{00}\rangle_E - \alpha_{11}^{01}|\Phi^-\rangle_{AB}|e_{11}^{01}\rangle_E - \alpha_{11}^{10}|\Psi^+\rangle_{AB}|e_{11}^{10}\rangle_E - \alpha_{11}^{11}|\Psi^-\rangle_{AB}|e_{11}^{11}\rangle_E \end{pmatrix}$$

$$= \frac{1}{2} \begin{bmatrix} |\Phi^+\rangle_{AB} \left( \alpha_{00}^{00} |e_{00}^{00}\rangle_E - \alpha_{01}^{00} |e_{01}^{00}\rangle_E + \alpha_{10}^{00} |e_{10}^{00}\rangle_E - \alpha_{11}^{00} |e_{11}^{00}\rangle_E \right) \\ + |\Phi^-\rangle_{AB} \left( \alpha_{00}^{01} |e_{00}^{01}\rangle_E - \alpha_{01}^{01} |e_{01}^{01}\rangle_E + \alpha_{10}^{01} |e_{10}^{01}\rangle_E - \alpha_{11}^{01} |e_{11}^{01}\rangle_E \right) \\ + |\Psi^+\rangle_{AB} \left( \alpha_{00}^{10} |e_{00}^{10}\rangle_E - \alpha_{01}^{10} |e_{01}^{10}\rangle_E + \alpha_{10}^{10} |e_{10}^{10}\rangle_E - \alpha_{11}^{10} |e_{11}^{10}\rangle_E \right) \\ + |\Psi^-\rangle_{AB} \left( \alpha_{00}^{11} |e_{00}^{11}\rangle_E - \alpha_{01}^{11} |e_{01}^{11}\rangle_E + \alpha_{10}^{11} |e_{10}^{11}\rangle_E - \alpha_{11}^{11} |e_{11}^{11}\rangle_E \right) \end{bmatrix} \quad (6)$$

Case 7:

$$U_E \left( |-\rangle_A \otimes |+\rangle_B \otimes |E_i\rangle_E \right)$$

$$= \frac{1}{2} U_E \left( |00E\rangle_{ABE} + |01E\rangle_{ABE} - |10E\rangle_{ABE} - |11E\rangle_{ABE} \right)$$

$$= \frac{1}{2} \begin{pmatrix} \alpha_{00}^{00} |\Phi^+\rangle_{AB} |e_{00}^{00}\rangle_E + \alpha_{00}^{01} |\Phi^-\rangle_{AB} |e_{00}^{01}\rangle_E + \alpha_{00}^{10} |\Psi^+\rangle_{AB} |e_{00}^{10}\rangle_E + \alpha_{00}^{11} |\Psi^-\rangle_{AB} |e_{00}^{11}\rangle_E \\ + \alpha_{01}^{00} |\Phi^+\rangle_{AB} |e_{01}^{00}\rangle_E + \alpha_{01}^{01} |\Phi^-\rangle_{AB} |e_{01}^{01}\rangle_E + \alpha_{01}^{10} |\Psi^+\rangle_{AB} |e_{01}^{10}\rangle_E + \alpha_{01}^{11} |\Psi^-\rangle_{AB} |e_{01}^{11}\rangle_E \\ - \alpha_{10}^{00} |\Phi^+\rangle_{AB} |e_{10}^{00}\rangle_E - \alpha_{10}^{01} |\Phi^-\rangle_{AB} |e_{10}^{01}\rangle_E - \alpha_{10}^{10} |\Psi^+\rangle_{AB} |e_{10}^{10}\rangle_E - \alpha_{10}^{11} |\Psi^-\rangle_{AB} |e_{10}^{11}\rangle_E \\ - \alpha_{11}^{00} |\Phi^+\rangle_{AB} |e_{11}^{00}\rangle_E - \alpha_{11}^{01} |\Phi^-\rangle_{AB} |e_{11}^{01}\rangle_E - \alpha_{11}^{10} |\Psi^+\rangle_{AB} |e_{11}^{10}\rangle_E - \alpha_{11}^{11} |\Psi^-\rangle_{AB} |e_{11}^{11}\rangle_E \end{pmatrix}$$

$$= \frac{1}{2} \begin{bmatrix} |\Phi^+\rangle_{AB} \left( \alpha_{00}^{00} |e_{00}^{00}\rangle_E + \alpha_{01}^{00} |e_{01}^{00}\rangle_E - \alpha_{10}^{00} |e_{10}^{00}\rangle_E - \alpha_{11}^{00} |e_{11}^{00}\rangle_E \right) \\ + |\Phi^-\rangle_{AB} \left( \alpha_{00}^{01} |e_{00}^{01}\rangle_E + \alpha_{01}^{01} |e_{01}^{01}\rangle_E - \alpha_{10}^{01} |e_{10}^{01}\rangle_E - \alpha_{11}^{01} |e_{11}^{01}\rangle_E \right) \\ + |\Psi^+\rangle_{AB} \left( \alpha_{00}^{10} |e_{00}^{10}\rangle_E + \alpha_{01}^{10} |e_{01}^{10}\rangle_E - \alpha_{10}^{10} |e_{10}^{10}\rangle_E - \alpha_{11}^{10} |e_{11}^{10}\rangle_E \right) \\ + |\Psi^-\rangle_{AB} \left( \alpha_{00}^{11} |e_{00}^{11}\rangle_E + \alpha_{01}^{11} |e_{01}^{11}\rangle_E - \alpha_{10}^{11} |e_{10}^{11}\rangle_E - \alpha_{11}^{11} |e_{11}^{11}\rangle_E \right) \end{bmatrix} \quad (7)$$

Case 8:

$$U_E \left( |-\rangle_A \otimes |-\rangle_B \otimes |E_i\rangle_E \right)$$

$$= \frac{1}{2} U_E \left( |00E\rangle_{ABE} - |01E\rangle_{ABE} - |10E\rangle_{ABE} + |11E\rangle_{ABE} \right)$$

$$= \frac{1}{2} \begin{pmatrix} \alpha_{00}^{00} |\Phi^+\rangle_{AB} |e_{00}^{00}\rangle_E + \alpha_{00}^{01} |\Phi^-\rangle_{AB} |e_{00}^{01}\rangle_E + \alpha_{00}^{10} |\Psi^+\rangle_{AB} |e_{00}^{10}\rangle_E + \alpha_{00}^{11} |\Psi^-\rangle_{AB} |e_{00}^{11}\rangle_E \\ - \alpha_{01}^{00} |\Phi^+\rangle_{AB} |e_{01}^{00}\rangle_E - \alpha_{01}^{01} |\Phi^-\rangle_{AB} |e_{01}^{01}\rangle_E - \alpha_{01}^{10} |\Psi^+\rangle_{AB} |e_{01}^{10}\rangle_E - \alpha_{01}^{11} |\Psi^-\rangle_{AB} |e_{01}^{11}\rangle_E \\ - \alpha_{10}^{00} |\Phi^+\rangle_{AB} |e_{10}^{00}\rangle_E - \alpha_{10}^{01} |\Phi^-\rangle_{AB} |e_{10}^{01}\rangle_E - \alpha_{10}^{10} |\Psi^+\rangle_{AB} |e_{10}^{10}\rangle_E - \alpha_{10}^{11} |\Psi^-\rangle_{AB} |e_{10}^{11}\rangle_E \\ + \alpha_{11}^{00} |\Phi^+\rangle_{AB} |e_{11}^{00}\rangle_E + \alpha_{11}^{01} |\Phi^-\rangle_{AB} |e_{11}^{01}\rangle_E + \alpha_{11}^{10} |\Psi^+\rangle_{AB} |e_{11}^{10}\rangle_E + \alpha_{11}^{11} |\Psi^-\rangle_{AB} |e_{11}^{11}\rangle_E \end{pmatrix}$$

$$= \frac{1}{2} \begin{bmatrix} |\Phi^+\rangle_{AB} \left( \alpha_{00}^{00} |e_{00}^{00}\rangle_E - \alpha_{01}^{00} |e_{01}^{00}\rangle_E - \alpha_{10}^{00} |e_{10}^{00}\rangle_E + \alpha_{11}^{00} |e_{11}^{00}\rangle_E \right) \\ + |\Phi^-\rangle_{AB} \left( \alpha_{00}^{01} |e_{00}^{01}\rangle_E - \alpha_{01}^{01} |e_{01}^{01}\rangle_E - \alpha_{10}^{01} |e_{10}^{01}\rangle_E + \alpha_{11}^{01} |e_{11}^{01}\rangle_E \right) \\ + |\Psi^+\rangle_{AB} \left( \alpha_{00}^{10} |e_{00}^{10}\rangle_E - \alpha_{01}^{10} |e_{01}^{10}\rangle_E - \alpha_{10}^{10} |e_{10}^{10}\rangle_E + \alpha_{11}^{10} |e_{11}^{10}\rangle_E \right) \\ + |\Psi^-\rangle_{AB} \left( \alpha_{00}^{11} |e_{00}^{11}\rangle_E - \alpha_{01}^{11} |e_{01}^{11}\rangle_E - \alpha_{10}^{11} |e_{10}^{11}\rangle_E + \alpha_{11}^{11} |e_{11}^{11}\rangle_E \right) \end{bmatrix} \quad (8)$$

In equations (1)-(8), $\sum_j |\alpha_i^j|^2 = 0$. Subsequently, TP uses Bell measurement to measure each qubit pair of $\{q_A^{ri}, q_B^{ri}\}$ and announces the measurement results. In Step 4 and Step 5, Alice and Bob use the measurement results to check whether there is an eavesdropper during the particle transmission processes. According to Table 1, to avoid being detected by the participants, TP must set all the parameters in the above equations (1)-(8) to meet the following conditions.

$$\alpha_{00}^{10} |e_{00}^{10}\rangle_E = \vec{0}$$
$$\alpha_{00}^{11} |e_{00}^{11}\rangle_E = \vec{0}$$
$$\alpha_{01}^{00} |e_{01}^{00}\rangle_E = \vec{0}$$
$$\alpha_{01}^{01} |e_{01}^{01}\rangle_E = \vec{0}$$
$$\alpha_{10}^{00} |e_{10}^{00}\rangle_E = \vec{0}$$
$$\alpha_{10}^{01} |e_{10}^{01}\rangle_E = \vec{0}$$
$$\alpha_{11}^{10} |e_{11}^{10}\rangle_E = \vec{0}$$
$$\alpha_{11}^{11} |e_{11}^{11}\rangle_E = \vec{0}$$

$$\alpha_{00}^{01} |e_{00}^{01}\rangle_E + \alpha_{01}^{01} |e_{01}^{01}\rangle_E + \alpha_{10}^{01} |e_{10}^{01}\rangle_E + \alpha_{11}^{01} |e_{11}^{01}\rangle_E = \vec{0}$$
$$\alpha_{00}^{11} |e_{00}^{11}\rangle_E + \alpha_{01}^{11} |e_{01}^{11}\rangle_E + \alpha_{10}^{11} |e_{10}^{11}\rangle_E + \alpha_{11}^{11} |e_{11}^{11}\rangle_E = \vec{0}$$
$$\alpha_{00}^{00} |e_{00}^{00}\rangle_E - \alpha_{01}^{00} |e_{01}^{00}\rangle_E + \alpha_{10}^{00} |e_{10}^{00}\rangle_E - \alpha_{11}^{00} |e_{11}^{00}\rangle_E = \vec{0}$$
$$\alpha_{00}^{10} |e_{00}^{10}\rangle_E - \alpha_{01}^{10} |e_{01}^{10}\rangle_E + \alpha_{10}^{10} |e_{10}^{10}\rangle_E - \alpha_{11}^{10} |e_{11}^{10}\rangle_E = \vec{0}$$
$$\alpha_{00}^{00} |e_{00}^{00}\rangle_E + \alpha_{01}^{00} |e_{01}^{00}\rangle_E - \alpha_{10}^{00} |e_{10}^{00}\rangle_E - \alpha_{11}^{00} |e_{11}^{00}\rangle_E = \vec{0}$$
$$\alpha_{00}^{10} |e_{00}^{10}\rangle_E + \alpha_{01}^{10} |e_{01}^{10}\rangle_E - \alpha_{10}^{10} |e_{10}^{10}\rangle_E - \alpha_{11}^{10} |e_{11}^{10}\rangle_E = \vec{0}$$

$$\alpha_{00}^{01}\left|e_{00}^{01}\right\rangle_E - \alpha_{01}^{01}\left|e_{01}^{01}\right\rangle_E - \alpha_{10}^{01}\left|e_{10}^{01}\right\rangle_E + \alpha_{11}^{01}\left|e_{11}^{01}\right\rangle_E = \vec{0}$$

$$\alpha_{00}^{11}\left|e_{00}^{11}\right\rangle_E - \alpha_{01}^{11}\left|e_{01}^{11}\right\rangle_E - \alpha_{10}^{11}\left|e_{10}^{11}\right\rangle_E + \alpha_{11}^{11}\left|e_{11}^{11}\right\rangle_E = \vec{0}$$

These above conditions can be simplified into the following conditions:

$\alpha_{00}^{10}\left|e_{00}^{10}\right\rangle_E = \vec{0}$ ; $\alpha_{00}^{11}\left|e_{00}^{11}\right\rangle_E = \vec{0}$ ; $\alpha_{01}^{00}\left|e_{01}^{00}\right\rangle_E = \vec{0}$ ; $\alpha_{01}^{01}\left|e_{01}^{01}\right\rangle_E = \vec{0}$ ; $\alpha_{10}^{00}\left|e_{10}^{00}\right\rangle_E = \vec{0}$ ;

$\alpha_{10}^{01}\left|e_{10}^{01}\right\rangle_E = \vec{0}$ ; $\alpha_{11}^{10}\left|e_{11}^{10}\right\rangle_E = \vec{0}$ ; $\alpha_{11}^{11}\left|e_{11}^{11}\right\rangle_E = \vec{0}$ ; $\alpha_{00}^{01}\left|e_{00}^{01}\right\rangle_E = -\alpha_{11}^{01}\left|e_{11}^{01}\right\rangle_E$ ; $\alpha_{01}^{11}\left|e_{01}^{11}\right\rangle_E =$

$-\alpha_{10}^{11}\left|e_{10}^{11}\right\rangle_E$ ; $\alpha_{00}^{00}\left|e_{00}^{00}\right\rangle_E = \alpha_{11}^{00}\left|e_{11}^{00}\right\rangle_E$ ; $\alpha_{01}^{10}\left|e_{01}^{10}\right\rangle_E = \alpha_{10}^{10}\left|e_{10}^{10}\right\rangle_E$. With these conditions, the

equations (1)-(8) can be simplified into the following equation group.

$$\begin{cases} U_E\left(\left|0\right\rangle_A \otimes \left|0\right\rangle_B \otimes \left|E_i\right\rangle_E\right) = \alpha_{00}^{00}\left|\Phi^+\right\rangle_{AB}\left|e_{00}^{00}\right\rangle_E + \alpha_{00}^{01}\left|\Phi^-\right\rangle_{AB}\left|e_{00}^{01}\right\rangle_E \\ U_E\left(\left|0\right\rangle_A \otimes \left|1\right\rangle_B \otimes \left|E_i\right\rangle_E\right) = \alpha_{01}^{10}\left|\Psi^+\right\rangle_{AB}\left|e_{01}^{10}\right\rangle_E + \alpha_{01}^{11}\left|\Psi^-\right\rangle_{AB}\left|e_{01}^{11}\right\rangle_E \\ U_E\left(\left|1\right\rangle_A \otimes \left|0\right\rangle_B \otimes \left|E_i\right\rangle_E\right) = \alpha_{01}^{10}\left|\Psi^+\right\rangle_{AB}\left|e_{01}^{10}\right\rangle_E + \alpha_{01}^{11}\left|\Psi^-\right\rangle_{AB}\left|e_{01}^{11}\right\rangle_E \\ U_E\left(\left|1\right\rangle_A \otimes \left|1\right\rangle_B \otimes \left|E_i\right\rangle_E\right) = \alpha_{00}^{00}\left|\Phi^+\right\rangle_{AB}\left|e_{00}^{00}\right\rangle_E + \alpha_{00}^{01}\left|\Phi^-\right\rangle_{AB}\left|e_{00}^{01}\right\rangle_E \\ U_E\left(\left|+\right\rangle_A \otimes \left|+\right\rangle_B \otimes \left|E_i\right\rangle_E\right) = \alpha_{00}^{00}\left|\Phi^+\right\rangle_{AB}\left|e_{00}^{00}\right\rangle_E + \alpha_{01}^{10}\left|\Psi^+\right\rangle_{AB}\left|e_{01}^{10}\right\rangle_E \\ U_E\left(\left|+\right\rangle_A \otimes \left|-\right\rangle_B \otimes \left|E_i\right\rangle_E\right) = \alpha_{00}^{01}\left|\Phi^-\right\rangle_{AB}\left|e_{00}^{01}\right\rangle_E - \alpha_{01}^{11}\left|\Psi^-\right\rangle_{AB}\left|e_{01}^{11}\right\rangle_E \\ U_E\left(\left|-\right\rangle_A \otimes \left|+\right\rangle_B \otimes \left|E_i\right\rangle_E\right) = \alpha_{00}^{01}\left|\Phi^-\right\rangle_{AB}\left|e_{00}^{01}\right\rangle_E + \alpha_{01}^{11}\left|\Psi^-\right\rangle_{AB}\left|e_{01}^{11}\right\rangle_E \\ U_E\left(\left|-\right\rangle_A \otimes \left|-\right\rangle_B \otimes \left|E_i\right\rangle_E\right) = \alpha_{00}^{00}\left|\Phi^+\right\rangle_{AB}\left|e_{00}^{00}\right\rangle_E - \alpha_{01}^{10}\left|\Psi^+\right\rangle_{AB}\left|e_{01}^{10}\right\rangle_E \end{cases} \quad (9)$$

The equation group (9) is the result of the collective attack on the proposed LAQKD protocol 1. TP tries to use the measurement result of his/her probe quantum system $E$ in equation group (9) to obtain the final shared key $K$ and the pre-shared key $K_1$. For easily understanding of the collective attack result, we re-express the equation group (9) and Table 1 into Table 4.

Table 4. Collective attack on the proposed LAQKD protocol 1

| MR of $E$ | MR of $\{Q'_A, Q'_B\}$ | $\{Q'_A, Q'_B\}$ | $R_A$ | $K_1$ |
| --- | --- | --- | --- | --- |
| $\left|e_{00}^{00}\right\rangle_E$ | $\left|\Phi^+\right\rangle_{AB}$ | $\{\left|0\right\rangle_A, \left|0\right\rangle_B\}$ | 0 | 0 |
| | | $\{\left|1\right\rangle_A, \left|1\right\rangle_B\}$ | 1 | 0 |

| | | $\{\|+\rangle_A, \|+\rangle_B\}$ | 0 | 1 |
| | | $\{\|-\rangle_A, \|-\rangle_B\}$ | 1 | 1 |
| $\|e_{00}^{01}\rangle_E$ | $\|\Phi^-\rangle_{AB}$ | $\{\|0\rangle_A, \|0\rangle_B\}$ | 0 | 0 |
| | | $\{\|1\rangle_A, \|1\rangle_B\}$ | 1 | 0 |
| | | $\{\|+\rangle_A, \|-\rangle_B\}$ | 0 | 1 |
| | | $\{\|-\rangle_A, \|+\rangle_B\}$ | 1 | 1 |
| $\|e_{01}^{10}\rangle_E$ | $\|\Psi^+\rangle_{AB}$ | $\{\|0\rangle_A, \|1\rangle_B\}$ | 0 | 0 |
| | | $\{\|1\rangle_A, \|0\rangle_B\}$ | 1 | 0 |
| | | $\{\|+\rangle_A, \|+\rangle_B\}$ | 0 | 1 |
| | | $\{\|-\rangle_A, \|-\rangle_B\}$ | 1 | 1 |
| $\|e_{01}^{11}\rangle_E$ | $\|\Psi^-\rangle_{AB}$ | $\{\|0\rangle_A, \|1\rangle_B\}$ | 0 | 0 |
| | | $\{\|1\rangle_A, \|0\rangle_B\}$ | 1 | 0 |
| | | $\{\|+\rangle_A, \|-\rangle_B\}$ | 0 | 1 |
| | | $\{\|-\rangle_A, \|+\rangle_B\}$ | 1 | 1 |

In Table 4, the 'MR' is the abbreviation of the measurement result. According to Table 4, we can find that no matter which measurement result of $E$ is obtained by TP, he/she can obtain nothing of $R_A$ and $K_1$. For example, if the measurement result of $E$ is $\|e_{00}^{00}\rangle_E$, the corresponding $(R_A, K_1)$ can be any one of $\{(0,0),(0,1),(1,0),(1,1)\}$. Moreover, $R_A$ is used for deriving the final shared key $K$.
20

Hence, this collective attack analysis result shows that any eavesdropper cannot obtain the final shared key $K$ or the pre-shared key $K_1$ by performing a collective attack without being detected. Theorem 1 is proved.

**Theorem 2:** In the proposed LAQKD protocol 2, no eavesdropper can obtain the final shared key $K$ or the pre-shared keys $K_1$ and $K_2$ by the collective attack without being detected.

**Proof:** Assume TP is an eavesdropper. In Step 1', TP generates arbitrary quantum systems to form $\{Q_A, Q_B\}$ instead. Then he/she generates his/her probe quantum systems $E=\{e^1=|E_1\rangle, e^2=|E_2\rangle, \cdots, e^n=|E_n\rangle\}$ and performs a joint unitary operation $U_E$ on each pair of $\{q_A^i, q_B^i, e^i\}$ to entangle them. The $U_E(q_A^i, q_B^i, e^i)$ can be described as follows.

$$U_E(q_A'^i, q_B'^i, e^i) \\ = a_{00}|00\rangle_{AB}|e_{00}\rangle_E + a_{01}|01\rangle_{AB}|e_{01}\rangle_E + a_{10}|10\rangle_{AB}|e_{10}\rangle_E + a_{11}|11\rangle_{AB}|e_{11}\rangle_E \tag{10}$$

Here, $\sum_j |a_j|^2 = 0$.

Subsequently, TP sends $Q_A$ and $Q_B$ to Alice and Bob, respectively. In Step 2', Alice (Bob) performs unitary operations $I$ or $H$ on $Q_A$ ($Q_B$) to obtain $Q_A'$ ($Q_B'$) according to $K_1$. This part can be considered as one of the two different joint unitary operations $U_{AB}^1 = I_A \otimes I_B \otimes I_E$ or $U_{AB}^2 = H_A \otimes H_B \otimes I_E$ is performed on the whole quantum system $\{q_A^i, q_B^i, e^i\}$ as follows.

$$U_{AB}^1 U_E(q_A'^i, q_B'^i, e^i) \\ = a_{00}|00\rangle_{AB}|e_{00}\rangle_E + a_{01}|01\rangle_{AB}|e_{01}\rangle_E + a_{10}|10\rangle_{AB}|e_{10}\rangle_E + a_{11}|11\rangle_{AB}|e_{11}\rangle_E \tag{11}$$

$$U_{AB}^2 U_E \left( q_A^{\prime i}, q_B^{\prime i}, e^i \right)$$
$$= a_{00} |++\rangle_{AB} |e_{00}\rangle_E + a_{01} |+-\rangle_{AB} |e_{01}\rangle_E + a_{10} |-+\rangle_{AB} |e_{10}\rangle_E + a_{11} |--\rangle_{AB} |e_{11}\rangle_E \quad (12)$$

$$= \frac{1}{2} \begin{bmatrix} |00\rangle_{AB} \left( a_{00} |e_{00}\rangle_E + a_{01} |e_{01}\rangle_E + a_{10} |e_{10}\rangle_E + a_{11} |e_{11}\rangle_E \right) \\ + |01\rangle_{AB} \left( a_{00} |e_{00}\rangle_E - a_{01} |e_{01}\rangle_E + a_{10} |e_{10}\rangle_E - a_{11} |e_{11}\rangle_E \right) \\ + |10\rangle_{AB} \left( a_{00} |e_{00}\rangle_E + a_{01} |e_{01}\rangle_E - a_{10} |e_{10}\rangle_E - a_{11} |e_{11}\rangle_E \right) \\ + |11\rangle_{AB} \left( a_{00} |e_{00}\rangle_E - a_{01} |e_{01}\rangle_E - a_{10} |e_{10}\rangle_E + a_{11} |e_{11}\rangle_E \right) \end{bmatrix}$$

In Step 2'-Step 4', Alice and Bob use the measurement result of $\{Q_A', Q_B'\}$ to check whether there is an eavesdropper during the particle transmission processes. To avoid being detected by Alice and Bob, TP must set all the parameters in equations (10)-(12) to meet the following conditions: $a_{01} |e_{01}\rangle_E = \vec{0}$; $a_{10} |e_{10}\rangle_E = \vec{0}$; $a_{00} |e_{00}\rangle_E = -a_{11} |e_{11}\rangle_E$; $a_{00} = \pm \frac{1}{\sqrt{2}}$. In this situation, the equations (11) and (12) can be transmitted into the following equation group.

$$\begin{cases} U_{AB}^1 U_E \left( q_A^{\prime i}, q_B^{\prime i}, e^i \right) = \pm \frac{1}{\sqrt{2}} \left( |00\rangle_{AB} - |11\rangle_{AB} \right) \otimes |e_{00}\rangle_E \\ U_{AB}^2 U_E \left( q_A^{\prime i}, q_B^{\prime i}, e^i \right) = \pm \frac{1}{\sqrt{2}} \left( |01\rangle_{AB} + |10\rangle_{AB} \right) \otimes |e_{00}\rangle_E \end{cases} \quad (13)$$

According to the equation group (13), we can find that the measurement result of $E$ always is $|e_{00}\rangle_E$. That means no matter which unitary operation is performed by Alice and Bob or which the exact bit of $R_A$ is, TP's measurement result of $E$ will always be $|e_{00}\rangle_E$. Hence, he/she can obtain nothing from the collective attack. The Theorem 2 is proved.

**Theorem 3:** In the proposed LAQKD protocol 3, no eavesdropper can obtain the final shared key $K$ or the pre-shared key $K_1$ by the collective attack without being detected.

**Proof:** Assume TP is an eavesdropper. In Step 1*, TP does not generate $|0\rangle$ to be the initial states. He/she generates arbitrary quantum systems to form $\{Q_A, Q_B\}$ instead. Then he/she generates his/her probe quantum systems $E_1 = \{e^1 = |E_1\rangle, e^2 = |E_2\rangle, \cdots, e^{n+m} = |E_{n+m}\rangle\}$ and performs a joint unitary operation $U_{E1}$ on each pair of $\{q_A^i, q_B^i, e^i\}$ to entangle them. The $U_{E1}(q_A^i, q_B^i, e^i)$ can be described as follows.

$$U_{E1}(q_A^i, q_B^i, e^i) \\ = a_1 |\Phi^+\rangle_{AB} |e_1\rangle_E + a_2 |\Phi^-\rangle_{AB} |e_2\rangle_E + a_3 |\Psi^+\rangle_{AB} |e_3\rangle_E + a_4 |\Psi^-\rangle_{AB} |e_4\rangle_E \tag{14}$$

Here, $\sum_j |a_j|^2 = 0$.

Then, TP sends $Q_A$ and $Q_B$ to Alice and Bob, respectively. In Step 3*, Alice (Bob) performs $j (j \in \{0,1,2,3\})$ times unitary operation $H'$ on $Q_A$ ($Q_B$) to obtain $Q_A'$ ($Q_B'$). Here, this part can be considered as one of the eight different joint unitary operations $\{U_{AB}^1, U_{AB}^2, \cdots, U_{AB}^8\}$ is performed on the whole quantum system $\{q_A^i, q_B^i, e^i\}$. The $\{U_{AB}^1, U_{AB}^2, \cdots, U_{AB}^8\}$ can be described as follows.

$$\begin{aligned}
U_{AB}^1 &= I_A \otimes I_B \otimes I_E \\
U_{AB}^2 &= I_A \otimes (H')^2_B \otimes I_E \\
U_{AB}^3 &= (H')^2_A \otimes I_B \otimes I_E \\
U_{AB}^4 &= (H')^2_A \otimes (H')^2_B \otimes I_E \\
U_{AB}^5 &= (H')^1_A \otimes (H')^1_B \otimes I_E \\
U_{AB}^6 &= (H')^1_A \otimes (H')^3_B \otimes I_E \\
U_{AB}^7 &= (H')^3_A \otimes (H')^1_B \otimes I_E \\
U_{AB}^8 &= (H')^3_A \otimes (H')^3_B \otimes I_E
\end{aligned} \tag{15}$$

According to equations (14) and (15), we can obtain the following equations.

$$U_{AB}^1 U_{E1}(q_A^i, q_B^i, e^i) \\ = a_1 |\Phi^+\rangle_{AB} |e_1\rangle_E + a_2 |\Phi^-\rangle_{AB} |e_2\rangle_E + a_3 |\Psi^+\rangle_{AB} |e_3\rangle_E + a_4 |\Psi^-\rangle_{AB} |e_4\rangle_E \tag{16}$$

$$U_{AB}^{2}U_{E1}\left(q_{A}^{i},q_{B}^{i},e^{i}\right)$$
$$=a_{1}\left|\Psi^{-}\right\rangle_{AB}\left|e_{1}\right\rangle_{E}+a_{2}\left|\Psi^{+}\right\rangle_{AB}\left|e_{2}\right\rangle_{E}-a_{3}\left|\Phi^{-}\right\rangle_{AB}\left|e_{3}\right\rangle_{E}-a_{4}\left|\Phi^{+}\right\rangle_{AB}\left|e_{4}\right\rangle_{E} \quad (17)$$

$$U_{AB}^{3}U_{E1}\left(q_{A}^{i},q_{B}^{i},e^{i}\right)$$
$$=-a_{1}\left|\Psi^{-}\right\rangle_{AB}\left|e_{1}\right\rangle_{E}+a_{2}\left|\Psi^{+}\right\rangle_{AB}\left|e_{2}\right\rangle_{E}-a_{3}\left|\Phi^{-}\right\rangle_{AB}\left|e_{3}\right\rangle_{E}+a_{4}\left|\Phi^{+}\right\rangle_{AB}\left|e_{4}\right\rangle_{E} \quad (18)$$

$$U_{AB}^{4}U_{E1}\left(q_{A}^{i},q_{B}^{i},e^{i}\right)$$
$$=a_{1}\left|\Phi^{+}\right\rangle_{AB}\left|e_{1}\right\rangle_{E}-a_{2}\left|\Phi^{-}\right\rangle_{AB}\left|e_{2}\right\rangle_{E}-a_{3}\left|\Psi^{+}\right\rangle_{AB}\left|e_{3}\right\rangle_{E}+a_{4}\left|\Psi^{-}\right\rangle_{AB}\left|e_{4}\right\rangle_{E} \quad (19)$$

$$U_{AB}^{5}U_{E1}\left(q_{A}^{i},q_{B}^{i},e^{i}\right)$$
$$=a_{1}\left|\Phi^{+}\right\rangle_{AB}\left|e_{1}\right\rangle_{E}+a_{2}\left|\Psi^{+}\right\rangle_{AB}\left|e_{2}\right\rangle_{E}-a_{3}\left|\Phi^{-}\right\rangle_{AB}\left|e_{3}\right\rangle_{E}+a_{4}\left|\Psi^{-}\right\rangle_{AB}\left|e_{4}\right\rangle_{E} \quad (20)$$

$$U_{AB}^{6}U_{E1}\left(q_{A}^{i},q_{B}^{i},e^{i}\right)$$
$$=a_{1}\left|\Psi^{-}\right\rangle_{AB}\left|e_{1}\right\rangle_{E}-a_{2}\left|\Phi^{-}\right\rangle_{AB}\left|e_{2}\right\rangle_{E}-a_{3}\left|\Psi^{+}\right\rangle_{AB}\left|e_{3}\right\rangle_{E}-a_{4}\left|\Phi^{+}\right\rangle_{AB}\left|e_{4}\right\rangle_{E} \quad (21)$$

$$U_{AB}^{7}U_{E1}\left(q_{A}^{i},q_{B}^{i},e^{i}\right)$$
$$=-a_{1}\left|\Psi^{-}\right\rangle_{AB}\left|e_{1}\right\rangle_{E}-a_{2}\left|\Phi^{-}\right\rangle_{AB}\left|e_{2}\right\rangle_{E}-a_{3}\left|\Psi^{+}\right\rangle_{AB}\left|e_{3}\right\rangle_{E}+a_{4}\left|\Phi^{+}\right\rangle_{AB}\left|e_{4}\right\rangle_{E} \quad (22)$$

$$U_{AB}^{8}U_{E1}\left(q_{A}^{i},q_{B}^{i},e^{i}\right)$$
$$=a_{1}\left|\Phi^{+}\right\rangle_{AB}\left|e_{1}\right\rangle_{E}-a_{2}\left|\Psi^{+}\right\rangle_{AB}\left|e_{2}\right\rangle_{E}+a_{3}\left|\Phi^{-}\right\rangle_{AB}\left|e_{3}\right\rangle_{E}+a_{4}\left|\Psi^{-}\right\rangle_{AB}\left|e_{4}\right\rangle_{E} \quad (23)$$

Then, Alice and Bob send $Q_{A}'$ and $Q_{B}'$ to TP, respectively. Upon receiving $Q_{A}'$ and $Q_{B}'$, TP performs another joint unitary operation $U_{E2}$ on each pair of $\{q_{A}'^{i}, q_{B}'^{i}, e^{i}\}$. The $U_{E2}$ can be described as follows.

$$U_{E2}\left(\left|x\right\rangle_{AB}\otimes\left|e_{y}\right\rangle_{E}\right)$$
$$=b_{xy}^{1}\left|\Phi^{+}\right\rangle_{AB}\left|e_{xy}^{1}\right\rangle_{E}+b_{xy}^{2}\left|\Phi^{-}\right\rangle_{AB}\left|e_{xy}^{2}\right\rangle_{E}+b_{xy}^{3}\left|\Psi^{+}\right\rangle_{AB}\left|e_{xy}^{3}\right\rangle_{E}+b_{xy}^{4}\left|\Psi^{-}\right\rangle_{AB}\left|e_{xy}^{4}\right\rangle_{E} \quad (24)$$

Here, $\sum_{j}\left|b_{xy}^{j}\right|^{2}=0$. Moreover, assume that '$\Phi^{+}$'='1', '$\Phi^{-}$'='2', '$\Psi^{+}$'='3' and '$\Psi^{-}$'='4', then $x,y\in\{1,2,3,4\}$. According to the equations (16)-(24), we can obtain the following equations.

$$U_{E2}U_{AB}^1U_{E1}\left(q_A^i, q_B^i, e^i\right)$$

$$=U_{E2}\left(a_1\left|\Phi^+\right\rangle_{AB}\left|e_1\right\rangle_E + a_2\left|\Phi^-\right\rangle_{AB}\left|e_2\right\rangle_E + a_3\left|\Psi^+\right\rangle_{AB}\left|e_3\right\rangle_E + a_4\left|\Psi^-\right\rangle_{AB}\left|e_4\right\rangle_E\right)$$

$$=a_1\left(b_{11}^1\left|\Phi^+\right\rangle_{AB}\left|e_{11}^1\right\rangle_E + b_{11}^2\left|\Phi^-\right\rangle_{AB}\left|e_{11}^2\right\rangle_E + b_{11}^3\left|\Psi^+\right\rangle_{AB}\left|e_{11}^3\right\rangle_E + b_{11}^4\left|\Psi^-\right\rangle_{AB}\left|e_{11}^4\right\rangle_E\right)$$

$$+a_2\left(b_{22}^1\left|\Phi^+\right\rangle_{AB}\left|e_{22}^1\right\rangle_E + b_{22}^2\left|\Phi^-\right\rangle_{AB}\left|e_{22}^2\right\rangle_E + b_{22}^3\left|\Psi^+\right\rangle_{AB}\left|e_{22}^3\right\rangle_E + b_{22}^4\left|\Psi^-\right\rangle_{AB}\left|e_{22}^4\right\rangle_E\right)$$

$$+a_3\left(b_{33}^1\left|\Phi^+\right\rangle_{AB}\left|e_{33}^1\right\rangle_E + b_{33}^2\left|\Phi^-\right\rangle_{AB}\left|e_{33}^2\right\rangle_E + b_{33}^3\left|\Psi^+\right\rangle_{AB}\left|e_{33}^3\right\rangle_E + b_{33}^4\left|\Psi^-\right\rangle_{AB}\left|e_{33}^4\right\rangle_E\right)$$

$$+a_4\left(b_{44}^1\left|\Phi^+\right\rangle_{AB}\left|e_{44}^1\right\rangle_E + b_{44}^2\left|\Phi^-\right\rangle_{AB}\left|e_{44}^2\right\rangle_E + b_{44}^3\left|\Psi^+\right\rangle_{AB}\left|e_{44}^3\right\rangle_E + b_{44}^4\left|\Psi^-\right\rangle_{AB}\left|e_{44}^4\right\rangle_E\right)$$

$$=\left|\Phi^+\right\rangle_{AB}\otimes\left(a_1b_{11}^1\left|e_{11}^1\right\rangle_E + a_2b_{22}^1\left|e_{22}^1\right\rangle_E + a_3b_{33}^1\left|e_{33}^1\right\rangle_E + a_4b_{44}^1\left|e_{44}^1\right\rangle_E\right)$$

$$+\left|\Phi^-\right\rangle_{AB}\otimes\left(a_1b_{11}^2\left|e_{11}^2\right\rangle_E + a_2b_{22}^2\left|e_{22}^2\right\rangle_E + a_3b_{33}^2\left|e_{33}^2\right\rangle_E + a_4b_{44}^2\left|e_{44}^2\right\rangle_E\right)$$

$$+\left|\Psi^+\right\rangle_{AB}\otimes\left(a_1b_{11}^3\left|e_{11}^3\right\rangle_E + a_2b_{22}^3\left|e_{22}^3\right\rangle_E + a_3b_{33}^3\left|e_{33}^3\right\rangle_E + a_4b_{44}^3\left|e_{44}^3\right\rangle_E\right)$$

$$+\left|\Psi^-\right\rangle_{AB}\otimes\left(a_1b_{11}^4\left|e_{11}^4\right\rangle_E + a_2b_{22}^4\left|e_{22}^4\right\rangle_E + a_3b_{33}^4\left|e_{33}^4\right\rangle_E + a_4b_{44}^4\left|e_{44}^4\right\rangle_E\right)$$

(25)

$$U_{E2}U_{AB}^2U_{E1}\left(q_A^i, q_B^i, e^i\right)$$

$$=U_{E2}\left(a_1\left|\Psi^-\right\rangle_{AB}\left|e_1\right\rangle_E + a_2\left|\Psi^+\right\rangle_{AB}\left|e_2\right\rangle_E - a_3\left|\Phi^-\right\rangle_{AB}\left|e_3\right\rangle_E - a_4\left|\Phi^+\right\rangle_{AB}\left|e_4\right\rangle_E\right)$$

$$=\left|\Phi^+\right\rangle_{AB}\otimes\left(a_1b_{41}^1\left|e_{41}^1\right\rangle_E + a_2b_{32}^1\left|e_{32}^1\right\rangle_E - a_3b_{23}^1\left|e_{23}^1\right\rangle_E - a_4b_{14}^1\left|e_{14}^1\right\rangle_E\right)$$

$$+\left|\Phi^-\right\rangle_{AB}\otimes\left(a_1b_{41}^2\left|e_{41}^2\right\rangle_E + a_2b_{32}^2\left|e_{32}^2\right\rangle_E - a_3b_{23}^2\left|e_{23}^2\right\rangle_E - a_4b_{14}^2\left|e_{14}^2\right\rangle_E\right)$$

$$+\left|\Psi^+\right\rangle_{AB}\otimes\left(a_1b_{41}^3\left|e_{41}^3\right\rangle_E + a_2b_{32}^3\left|e_{32}^3\right\rangle_E - a_3b_{23}^3\left|e_{23}^3\right\rangle_E - a_4b_{14}^3\left|e_{14}^3\right\rangle_E\right)$$

$$+\left|\Psi^-\right\rangle_{AB}\otimes\left(a_1b_{41}^4\left|e_{41}^4\right\rangle_E + a_2b_{32}^4\left|e_{32}^4\right\rangle_E - a_3b_{23}^4\left|e_{23}^4\right\rangle_E - a_4b_{14}^4\left|e_{14}^4\right\rangle_E\right)$$

(26)

$$U_{E2}U_{AB}^3U_{E1}\left(q_A^i, q_B^i, e^i\right)$$

$$=U_{E2}\left(-a_1\left|\Psi^-\right\rangle_{AB}\left|e_1\right\rangle_E + a_2\left|\Psi^+\right\rangle_{AB}\left|e_2\right\rangle_E - a_3\left|\Phi^-\right\rangle_{AB}\left|e_3\right\rangle_E + a_4\left|\Phi^+\right\rangle_{AB}\left|e_4\right\rangle_E\right)$$

$$=\left|\Phi^+\right\rangle_{AB}\otimes\left(-a_1b_{41}^1\left|e_{41}^1\right\rangle_E + a_2b_{32}^1\left|e_{32}^1\right\rangle_E - a_3b_{23}^1\left|e_{23}^1\right\rangle_E + a_4b_{14}^1\left|e_{14}^1\right\rangle_E\right)$$

$$+\left|\Phi^-\right\rangle_{AB}\otimes\left(-a_1b_{41}^2\left|e_{41}^2\right\rangle_E + a_2b_{32}^2\left|e_{32}^2\right\rangle_E - a_3b_{23}^2\left|e_{23}^2\right\rangle_E + a_4b_{14}^2\left|e_{14}^2\right\rangle_E\right)$$

$$+\left|\Psi^+\right\rangle_{AB}\otimes\left(-a_1b_{41}^3\left|e_{41}^3\right\rangle_E + a_2b_{32}^3\left|e_{32}^3\right\rangle_E - a_3b_{23}^3\left|e_{23}^3\right\rangle_E + a_4b_{14}^3\left|e_{14}^3\right\rangle_E\right)$$

$$+\left|\Psi^-\right\rangle_{AB}\otimes\left(-a_1b_{41}^4\left|e_{41}^4\right\rangle_E + a_2b_{32}^4\left|e_{32}^4\right\rangle_E - a_3b_{23}^4\left|e_{23}^4\right\rangle_E + a_4b_{14}^4\left|e_{14}^4\right\rangle_E\right)$$

(27)

$$U_{E2}U_{AB}^{4}U_{E1}\left(q_{A}^{i},q_{B}^{i},e^{i}\right)$$

$$=U_{E2}\left(a_{1}\left|\Phi^{+}\right\rangle_{AB}\left|e_{1}\right\rangle_{E}-a_{2}\left|\Phi^{-}\right\rangle_{AB}\left|e_{2}\right\rangle_{E}-a_{3}\left|\Psi^{+}\right\rangle_{AB}\left|e_{3}\right\rangle_{E}+a_{4}\left|\Psi^{-}\right\rangle_{AB}\left|e_{4}\right\rangle_{E}\right)$$

$$=\left|\Phi^{+}\right\rangle_{AB}\otimes\left(a_{1}b_{11}^{1}\left|e_{11}^{1}\right\rangle_{E}-a_{2}b_{22}^{1}\left|e_{22}^{1}\right\rangle_{E}-a_{3}b_{33}^{1}\left|e_{33}^{1}\right\rangle_{E}+a_{4}b_{44}^{1}\left|e_{44}^{1}\right\rangle_{E}\right)$$

$$+\left|\Phi^{-}\right\rangle_{AB}\otimes\left(a_{1}b_{11}^{2}\left|e_{11}^{2}\right\rangle_{E}-a_{2}b_{22}^{2}\left|e_{22}^{2}\right\rangle_{E}-a_{3}b_{33}^{2}\left|e_{33}^{2}\right\rangle_{E}+a_{4}b_{44}^{2}\left|e_{44}^{2}\right\rangle_{E}\right)$$

$$+\left|\Psi^{+}\right\rangle_{AB}\otimes\left(a_{1}b_{11}^{3}\left|e_{11}^{3}\right\rangle_{E}-a_{2}b_{22}^{3}\left|e_{22}^{3}\right\rangle_{E}-a_{3}b_{33}^{3}\left|e_{33}^{3}\right\rangle_{E}+a_{4}b_{44}^{3}\left|e_{44}^{3}\right\rangle_{E}\right)$$

$$+\left|\Psi^{-}\right\rangle_{AB}\otimes\left(a_{1}b_{11}^{4}\left|e_{11}^{4}\right\rangle_{E}-a_{2}b_{22}^{4}\left|e_{22}^{4}\right\rangle_{E}-a_{3}b_{33}^{4}\left|e_{33}^{4}\right\rangle_{E}+a_{4}b_{44}^{4}\left|e_{44}^{4}\right\rangle_{E}\right)$$

(28)

$$U_{E2}U_{AB}^{5}U_{E1}\left(q_{A}^{i},q_{B}^{i},e^{i}\right)$$

$$=U_{E2}\left(a_{1}\left|\Phi^{+}\right\rangle_{AB}\left|e_{1}\right\rangle_{E}+a_{2}\left|\Psi^{+}\right\rangle_{AB}\left|e_{2}\right\rangle_{E}-a_{3}\left|\Phi^{-}\right\rangle_{AB}\left|e_{3}\right\rangle_{E}+a_{4}\left|\Psi^{-}\right\rangle_{AB}\left|e_{4}\right\rangle_{E}\right)$$

$$=\left|\Phi^{+}\right\rangle_{AB}\otimes\left(a_{1}b_{11}^{1}\left|e_{11}^{1}\right\rangle_{E}+a_{2}b_{32}^{1}\left|e_{32}^{1}\right\rangle_{E}-a_{3}b_{23}^{1}\left|e_{23}^{1}\right\rangle_{E}+a_{4}b_{44}^{1}\left|e_{44}^{1}\right\rangle_{E}\right)$$

$$+\left|\Phi^{-}\right\rangle_{AB}\otimes\left(a_{1}b_{11}^{2}\left|e_{11}^{2}\right\rangle_{E}+a_{2}b_{32}^{2}\left|e_{32}^{2}\right\rangle_{E}-a_{3}b_{23}^{2}\left|e_{23}^{2}\right\rangle_{E}+a_{4}b_{44}^{2}\left|e_{44}^{2}\right\rangle_{E}\right)$$

$$+\left|\Psi^{+}\right\rangle_{AB}\otimes\left(a_{1}b_{11}^{3}\left|e_{11}^{3}\right\rangle_{E}+a_{2}b_{32}^{3}\left|e_{32}^{3}\right\rangle_{E}-a_{3}b_{23}^{3}\left|e_{23}^{3}\right\rangle_{E}+a_{4}b_{44}^{3}\left|e_{44}^{3}\right\rangle_{E}\right)$$

$$+\left|\Psi^{-}\right\rangle_{AB}\otimes\left(a_{1}b_{11}^{4}\left|e_{11}^{4}\right\rangle_{E}+a_{2}b_{32}^{4}\left|e_{32}^{4}\right\rangle_{E}-a_{3}b_{23}^{4}\left|e_{23}^{4}\right\rangle_{E}+a_{4}b_{44}^{4}\left|e_{44}^{4}\right\rangle_{E}\right)$$

(29)

$$U_{E2}U_{AB}^{6}U_{E1}\left(q_{A}^{i},q_{B}^{i},e^{i}\right)$$

$$=U_{E2}\left(a_{1}\left|\Psi^{-}\right\rangle_{AB}\left|e_{1}\right\rangle_{E}-a_{2}\left|\Phi^{-}\right\rangle_{AB}\left|e_{2}\right\rangle_{E}-a_{3}\left|\Psi^{+}\right\rangle_{AB}\left|e_{3}\right\rangle_{E}-a_{4}\left|\Phi^{+}\right\rangle_{AB}\left|e_{4}\right\rangle_{E}\right)$$

$$=\left|\Phi^{+}\right\rangle_{AB}\otimes\left(a_{1}b_{41}^{1}\left|e_{41}^{1}\right\rangle_{E}-a_{2}b_{22}^{1}\left|e_{22}^{1}\right\rangle_{E}-a_{3}b_{33}^{1}\left|e_{33}^{1}\right\rangle_{E}-a_{4}b_{14}^{1}\left|e_{14}^{1}\right\rangle_{E}\right)$$

$$+\left|\Phi^{-}\right\rangle_{AB}\otimes\left(a_{1}b_{41}^{2}\left|e_{41}^{2}\right\rangle_{E}-a_{2}b_{22}^{2}\left|e_{22}^{2}\right\rangle_{E}-a_{3}b_{33}^{2}\left|e_{33}^{2}\right\rangle_{E}-a_{4}b_{14}^{2}\left|e_{14}^{2}\right\rangle_{E}\right)$$

$$+\left|\Psi^{+}\right\rangle_{AB}\otimes\left(a_{1}b_{41}^{3}\left|e_{41}^{3}\right\rangle_{E}-a_{2}b_{22}^{3}\left|e_{22}^{3}\right\rangle_{E}-a_{3}b_{33}^{3}\left|e_{33}^{3}\right\rangle_{E}-a_{4}b_{14}^{3}\left|e_{14}^{3}\right\rangle_{E}\right)$$

$$+\left|\Psi^{-}\right\rangle_{AB}\otimes\left(a_{1}b_{41}^{4}\left|e_{41}^{4}\right\rangle_{E}-a_{2}b_{22}^{4}\left|e_{22}^{4}\right\rangle_{E}-a_{3}b_{33}^{4}\left|e_{33}^{4}\right\rangle_{E}-a_{4}b_{14}^{4}\left|e_{14}^{4}\right\rangle_{E}\right)$$

(30)

$$U_{E2}U_{AB}^{7}U_{E1}\left(q_{A}^{i},q_{B}^{i},e^{i}\right)$$

$$=U_{E2}\left(-a_{1}\left|\Psi^{-}\right\rangle_{AB}\left|e_{1}\right\rangle_{E}-a_{2}\left|\Phi^{-}\right\rangle_{AB}\left|e_{2}\right\rangle_{E}-a_{3}\left|\Psi^{+}\right\rangle_{AB}\left|e_{3}\right\rangle_{E}+a_{4}\left|\Phi^{+}\right\rangle_{AB}\left|e_{4}\right\rangle_{E}\right)$$

$$=\left|\Phi^{+}\right\rangle_{AB}\otimes\left(-a_{1}b_{41}^{1}\left|e_{41}^{1}\right\rangle_{E}-a_{2}b_{22}^{1}\left|e_{22}^{1}\right\rangle_{E}-a_{3}b_{33}^{1}\left|e_{33}^{1}\right\rangle_{E}+a_{4}b_{14}^{1}\left|e_{14}^{1}\right\rangle_{E}\right)$$

$$+\left|\Phi^{-}\right\rangle_{AB}\otimes\left(-a_{1}b_{41}^{2}\left|e_{41}^{2}\right\rangle_{E}-a_{2}b_{22}^{2}\left|e_{22}^{2}\right\rangle_{E}-a_{3}b_{33}^{2}\left|e_{33}^{2}\right\rangle_{E}+a_{4}b_{14}^{2}\left|e_{14}^{2}\right\rangle_{E}\right)$$

$$+\left|\Psi^{+}\right\rangle_{AB}\otimes\left(-a_{1}b_{41}^{3}\left|e_{41}^{3}\right\rangle_{E}-a_{2}b_{22}^{3}\left|e_{22}^{3}\right\rangle_{E}-a_{3}b_{33}^{3}\left|e_{33}^{3}\right\rangle_{E}+a_{4}b_{14}^{3}\left|e_{14}^{3}\right\rangle_{E}\right)$$

$$+\left|\Psi^{-}\right\rangle_{AB}\otimes\left(-a_{1}b_{41}^{4}\left|e_{41}^{4}\right\rangle_{E}-a_{2}b_{22}^{4}\left|e_{22}^{4}\right\rangle_{E}-a_{3}b_{33}^{4}\left|e_{33}^{4}\right\rangle_{E}+a_{4}b_{14}^{4}\left|e_{14}^{4}\right\rangle_{E}\right)$$

(31)

$$U_{E2}U_{AB}^{8}U_{E1}\left(q_{A}^{i},q_{B}^{i},e^{i}\right)$$

$$=U_{E2}\left(a_{1}\left|\Phi^{+}\right\rangle_{AB}\left|e_{1}\right\rangle_{E}-a_{2}\left|\Psi^{+}\right\rangle_{AB}\left|e_{2}\right\rangle_{E}+a_{3}\left|\Phi^{-}\right\rangle_{AB}\left|e_{3}\right\rangle_{E}+a_{4}\left|\Psi^{-}\right\rangle_{AB}\left|e_{4}\right\rangle_{E}\right)$$

$$=\left|\Phi^{+}\right\rangle_{AB}\otimes\left(a_{1}b_{11}^{1}\left|e_{11}^{1}\right\rangle_{E}-a_{2}b_{32}^{1}\left|e_{32}^{1}\right\rangle_{E}+a_{3}b_{23}^{1}\left|e_{23}^{1}\right\rangle_{E}+a_{4}b_{44}^{1}\left|e_{44}^{1}\right\rangle_{E}\right)$$

$$+\left|\Phi^{-}\right\rangle_{AB}\otimes\left(a_{1}b_{11}^{2}\left|e_{11}^{2}\right\rangle_{E}-a_{2}b_{32}^{2}\left|e_{32}^{2}\right\rangle_{E}+a_{3}b_{23}^{2}\left|e_{23}^{2}\right\rangle_{E}+a_{4}b_{44}^{2}\left|e_{44}^{2}\right\rangle_{E}\right) \quad (32)$$

$$+\left|\Psi^{+}\right\rangle_{AB}\otimes\left(a_{1}b_{11}^{3}\left|e_{11}^{3}\right\rangle_{E}-a_{2}b_{32}^{3}\left|e_{32}^{3}\right\rangle_{E}+a_{3}b_{23}^{3}\left|e_{23}^{3}\right\rangle_{E}+a_{4}b_{44}^{3}\left|e_{44}^{3}\right\rangle_{E}\right)$$

$$+\left|\Psi^{-}\right\rangle_{AB}\otimes\left(a_{1}b_{11}^{4}\left|e_{11}^{4}\right\rangle_{E}-a_{2}b_{32}^{4}\left|e_{32}^{4}\right\rangle_{E}+a_{3}b_{23}^{4}\left|e_{23}^{4}\right\rangle_{E}+a_{4}b_{44}^{4}\left|e_{44}^{4}\right\rangle_{E}\right)$$

Subsequently, TP performs Bell measurement on $\{Q_{A}',Q_{B}'\}$ and announces the measurement results. In Step 5* and Step 6*, Alice and Bob use the announced measurement results to check whether there is an eavesdropper during the qubit transmission processes. To avoid being detected by Alice and Bob, TP must set all the parameters in equations (25)-(32) to meet the following conditions:

$$a_{1}b_{11}^{3}\left|e_{11}^{3}\right\rangle_{E}+a_{2}b_{22}^{3}\left|e_{22}^{3}\right\rangle_{E}+a_{3}b_{33}^{3}\left|e_{33}^{3}\right\rangle_{E}+a_{4}b_{44}^{3}\left|e_{44}^{3}\right\rangle_{E}=\vec{0}$$

$$a_{1}b_{11}^{4}\left|e_{11}^{4}\right\rangle_{E}+a_{2}b_{22}^{4}\left|e_{22}^{4}\right\rangle_{E}+a_{3}b_{33}^{4}\left|e_{33}^{4}\right\rangle_{E}+a_{4}b_{44}^{4}\left|e_{44}^{4}\right\rangle_{E}=\vec{0}$$

$$a_{1}b_{41}^{1}\left|e_{41}^{1}\right\rangle_{E}+a_{2}b_{32}^{1}\left|e_{32}^{1}\right\rangle_{E}-a_{3}b_{23}^{1}\left|e_{23}^{1}\right\rangle_{E}-a_{4}b_{14}^{1}\left|e_{14}^{1}\right\rangle_{E}=\vec{0}$$

$$a_{1}b_{41}^{2}\left|e_{41}^{2}\right\rangle_{E}+a_{2}b_{32}^{2}\left|e_{32}^{2}\right\rangle_{E}-a_{3}b_{23}^{2}\left|e_{23}^{2}\right\rangle_{E}-a_{4}b_{14}^{2}\left|e_{14}^{2}\right\rangle_{E}=\vec{0}$$

$$-a_{1}b_{41}^{1}\left|e_{41}^{1}\right\rangle_{E}+a_{2}b_{32}^{1}\left|e_{32}^{1}\right\rangle_{E}-a_{3}b_{23}^{1}\left|e_{23}^{1}\right\rangle_{E}+a_{4}b_{14}^{1}\left|e_{14}^{1}\right\rangle_{E}=\vec{0}$$

$$-a_{1}b_{41}^{2}\left|e_{41}^{2}\right\rangle_{E}+a_{2}b_{32}^{2}\left|e_{32}^{2}\right\rangle_{E}-a_{3}b_{23}^{2}\left|e_{23}^{2}\right\rangle_{E}+a_{4}b_{14}^{2}\left|e_{14}^{2}\right\rangle_{E}=\vec{0}$$

$$a_{1}b_{11}^{3}\left|e_{11}^{3}\right\rangle_{E}-a_{2}b_{22}^{3}\left|e_{22}^{3}\right\rangle_{E}-a_{3}b_{33}^{3}\left|e_{33}^{3}\right\rangle_{E}+a_{4}b_{44}^{3}\left|e_{44}^{3}\right\rangle_{E}=\vec{0}$$

$$a_{1}b_{11}^{4}\left|e_{11}^{4}\right\rangle_{E}-a_{2}b_{22}^{4}\left|e_{22}^{4}\right\rangle_{E}-a_{3}b_{33}^{4}\left|e_{33}^{4}\right\rangle_{E}+a_{4}b_{44}^{4}\left|e_{44}^{4}\right\rangle_{E}=\vec{0}$$

$$a_{1}b_{11}^{2}\left|e_{11}^{2}\right\rangle_{E}+a_{2}b_{32}^{2}\left|e_{32}^{2}\right\rangle_{E}-a_{3}b_{23}^{2}\left|e_{23}^{2}\right\rangle_{E}+a_{4}b_{44}^{2}\left|e_{44}^{2}\right\rangle_{E}=\vec{0}$$

$$a_{1}b_{11}^{4}\left|e_{11}^{4}\right\rangle_{E}+a_{2}b_{32}^{4}\left|e_{32}^{4}\right\rangle_{E}-a_{3}b_{23}^{4}\left|e_{23}^{4}\right\rangle_{E}+a_{4}b_{44}^{4}\left|e_{44}^{4}\right\rangle_{E}=\vec{0}$$

$$a_{1}b_{41}^{1}\left|e_{41}^{1}\right\rangle_{E}-a_{2}b_{22}^{1}\left|e_{22}^{1}\right\rangle_{E}-a_{3}b_{33}^{1}\left|e_{33}^{1}\right\rangle_{E}-a_{4}b_{14}^{1}\left|e_{14}^{1}\right\rangle_{E}=\vec{0}$$

$$a_{1}b_{41}^{3}\left|e_{41}^{3}\right\rangle_{E}-a_{2}b_{22}^{3}\left|e_{22}^{3}\right\rangle_{E}-a_{3}b_{33}^{3}\left|e_{33}^{3}\right\rangle_{E}-a_{4}b_{14}^{3}\left|e_{14}^{3}\right\rangle_{E}=\vec{0}$$

$$-a_{1}b_{41}^{1}\left|e_{41}^{1}\right\rangle_{E}-a_{2}b_{22}^{1}\left|e_{22}^{1}\right\rangle_{E}-a_{3}b_{33}^{1}\left|e_{33}^{1}\right\rangle_{E}+a_{4}b_{14}^{1}\left|e_{14}^{1}\right\rangle_{E}=\vec{0}$$

$$-a_{1}b_{41}^{3}\left|e_{41}^{3}\right\rangle_{E}-a_{2}b_{22}^{3}\left|e_{22}^{3}\right\rangle_{E}-a_{3}b_{33}^{3}\left|e_{33}^{3}\right\rangle_{E}+a_{4}b_{14}^{3}\left|e_{14}^{3}\right\rangle_{E}=\vec{0}$$

$$a_1b_{11}^2\left|e_{11}^2\right\rangle_E - a_2b_{32}^2\left|e_{32}^2\right\rangle_E + a_3b_{23}^2\left|e_{23}^2\right\rangle_E + a_4b_{44}^2\left|e_{44}^2\right\rangle_E = \vec{0}$$

$$a_1b_{11}^4\left|e_{11}^4\right\rangle_E - a_2b_{32}^4\left|e_{32}^4\right\rangle_E + a_3b_{23}^4\left|e_{23}^4\right\rangle_E + a_4b_{44}^4\left|e_{44}^4\right\rangle_E = \vec{0}$$

The above conditions can be simplified into the following conditions:

$$\begin{cases} a_1b_{11}^3\left|e_{11}^3\right\rangle_E = -a_4b_{44}^3\left|e_{44}^3\right\rangle_E \\ a_1b_{11}^4\left|e_{11}^4\right\rangle_E = -a_4b_{44}^4\left|e_{44}^4\right\rangle_E \\ a_2b_{22}^3\left|e_{22}^3\right\rangle_E = -a_3b_{33}^3\left|e_{33}^3\right\rangle_E \\ a_2b_{22}^4\left|e_{22}^4\right\rangle_E = -a_3b_{33}^4\left|e_{33}^4\right\rangle_E \end{cases} \quad \begin{cases} a_1b_{41}^1\left|e_{41}^1\right\rangle_E = a_4b_{14}^1\left|e_{14}^1\right\rangle_E \\ a_1b_{41}^2\left|e_{41}^2\right\rangle_E = a_4b_{14}^2\left|e_{14}^2\right\rangle_E \\ a_2b_{32}^1\left|e_{32}^1\right\rangle_E = a_3b_{23}^1\left|e_{23}^1\right\rangle_E \\ a_2b_{32}^2\left|e_{32}^2\right\rangle_E = a_3b_{23}^2\left|e_{23}^2\right\rangle_E \end{cases}$$

$$\begin{cases} a_2b_{32}^2\left|e_{32}^2\right\rangle_E = a_3b_{23}^2\left|e_{23}^2\right\rangle_E \\ a_1b_{11}^4\left|e_{11}^4\right\rangle_E = -a_4b_{44}^4\left|e_{44}^4\right\rangle_E \\ a_1b_{11}^2\left|e_{11}^2\right\rangle_E = -a_4b_{44}^2\left|e_{44}^2\right\rangle_E \\ a_2b_{32}^4\left|e_{32}^4\right\rangle_E = a_3b_{23}^4\left|e_{23}^4\right\rangle_E \end{cases} \quad \begin{cases} a_1b_{41}^1\left|e_{41}^1\right\rangle_E = a_4b_{14}^1\left|e_{14}^1\right\rangle_E \\ a_1b_{41}^3\left|e_{41}^3\right\rangle_E = a_4b_{14}^3\left|e_{14}^3\right\rangle_E \\ a_2b_{22}^1\left|e_{22}^1\right\rangle_E = -a_3b_{33}^1\left|e_{33}^1\right\rangle_E \\ a_2b_{22}^3\left|e_{22}^3\right\rangle_E = -a_3b_{33}^3\left|e_{33}^3\right\rangle_E \end{cases}$$

According to these conditions, the equations (25)-(32) can be transformed into the following equations.

$$U_{E2}U_{AB}^1 U_{E1}\left(q_A^i, q_B^i, e^i\right) \\ = \left|\Phi^+\right\rangle_{AB} \otimes \left(a_1b_{11}^1\left|e_{11}^1\right\rangle_E + a_4b_{44}^1\left|e_{44}^1\right\rangle_E\right) + \left|\Phi^-\right\rangle_{AB} \otimes \left(a_2b_{22}^2\left|e_{22}^2\right\rangle_E + a_3b_{33}^2\left|e_{33}^2\right\rangle_E\right) \tag{33}$$

$$U_{E2}U_{AB}^2 U_{E1}\left(q_A^i, q_B^i, e^i\right) \\ = \left|\Psi^+\right\rangle_{AB} \otimes \left(a_2b_{32}^3\left|e_{32}^3\right\rangle_E - a_3b_{23}^3\left|e_{23}^3\right\rangle_E\right) + \left|\Psi^-\right\rangle_{AB} \otimes \left(a_1b_{41}^4\left|e_{41}^4\right\rangle_E - a_4b_{14}^4\left|e_{14}^4\right\rangle_E\right) \tag{34}$$

$$U_{E2}U_{AB}^3 U_{E1}\left(q_A^i, q_B^i, e^i\right) \\ = \left|\Psi^+\right\rangle_{AB} \otimes \left(a_2b_{32}^3\left|e_{32}^3\right\rangle_E - a_3b_{23}^3\left|e_{23}^3\right\rangle_E\right) - \left|\Psi^-\right\rangle_{AB} \otimes \left(a_1b_{41}^4\left|e_{41}^4\right\rangle_E - a_4b_{14}^4\left|e_{14}^4\right\rangle_E\right) \tag{35}$$

$$U_{E2}U_{AB}^4 U_{E1}\left(q_A^i, q_B^i, e^i\right) \\ = \left|\Phi^+\right\rangle_{AB} \otimes \left(a_1b_{11}^1\left|e_{11}^1\right\rangle_E + a_4b_{44}^1\left|e_{44}^1\right\rangle_E\right) - \left|\Phi^-\right\rangle_{AB} \otimes \left(a_2b_{22}^2\left|e_{22}^2\right\rangle_E + a_3b_{33}^2\left|e_{33}^2\right\rangle_E\right) \tag{36}$$

$$U_{E2}U_{AB}^5 U_{E1}\left(q_A^i, q_B^i, e^i\right) \\ = \left|\Phi^+\right\rangle_{AB} \otimes \left(a_1b_{11}^1\left|e_{11}^1\right\rangle_E + a_4b_{44}^1\left|e_{44}^1\right\rangle_E\right) + \left|\Psi^+\right\rangle_{AB} \otimes \left(a_2b_{32}^3\left|e_{32}^3\right\rangle_E - a_3b_{23}^3\left|e_{23}^3\right\rangle_E\right) \tag{37}$$

$$U_{E2}U_{AB}^6 U_{E1}\left(q_A^i, q_B^i, e^i\right) \\ = -\left|\Phi^-\right\rangle_{AB} \otimes \left(a_2b_{22}^2\left|e_{22}^2\right\rangle_E + a_3b_{33}^2\left|e_{33}^2\right\rangle_E\right) + \left|\Psi^-\right\rangle_{AB} \otimes \left(a_1b_{41}^4\left|e_{41}^4\right\rangle_E - a_4b_{14}^4\left|e_{14}^4\right\rangle_E\right) \tag{38}$$

$$U_{E2}U_{AB}^{7}U_{E1}(q_A^i, q_B^i, e^i)$$
$$= -|\Phi^-\rangle_{AB} \otimes (a_2 b_{22}^2 |e_{22}^2\rangle_E + a_3 b_{33}^2 |e_{33}^2\rangle_E) - |\Psi^-\rangle_{AB} \otimes (a_1 b_{41}^4 |e_{41}^4\rangle_E - a_4 b_{14}^4 |e_{14}^4\rangle_E) \quad (39)$$

$$U_{E2}U_{AB}^{8}U_{E1}(q_A^i, q_B^i, e^i)$$
$$= |\Phi^+\rangle_{AB} \otimes (a_1 b_{11}^1 |e_{11}^1\rangle_E + a_4 b_{44}^1 |e_{44}^1\rangle_E) - |\Psi^+\rangle_{AB} \otimes (a_2 b_{32}^3 |e_{32}^3\rangle_E - a_3 b_{23}^3 |e_{23}^3\rangle_E) \quad (40)$$

Assume that $|f_1\rangle_E = a_1 b_{11}^1 |e_{11}^1\rangle_E + a_4 b_{44}^1 |e_{44}^1\rangle_E$, $|f_2\rangle_E = a_2 b_{22}^2 |e_{22}^2\rangle_E + a_3 b_{33}^2 |e_{33}^2\rangle_E$, $|f_3\rangle_E = a_2 b_{32}^3 |e_{32}^3\rangle_E - a_3 b_{23}^3 |e_{23}^3\rangle_E$ and $|f_4\rangle_E = a_1 b_{41}^4 |e_{41}^4\rangle_E - a_4 b_{14}^4 |e_{14}^4\rangle_E$, then the equations (33)-(40) can be represented as the following equation group.

$$\begin{cases} U_{E2}U_{AB}^{1}U_{E1}(q_A^i, q_B^i, e^i) = |\Phi^+\rangle_{AB}|f_1\rangle_E + |\Phi^-\rangle_{AB}|f_2\rangle_E \\ U_{E2}U_{AB}^{2}U_{E1}(q_A^i, q_B^i, e^i) = |\Psi^+\rangle_{AB}|f_3\rangle_E + |\Psi^-\rangle_{AB}|f_4\rangle_E \\ U_{E2}U_{AB}^{3}U_{E1}(q_A^i, q_B^i, e^i) = |\Psi^+\rangle_{AB}|f_3\rangle_E - |\Psi^-\rangle_{AB}|f_4\rangle_E \\ U_{E2}U_{AB}^{4}U_{E1}(q_A^i, q_B^i, e^i) = |\Phi^+\rangle_{AB}|f_1\rangle_E - |\Phi^-\rangle_{AB}|f_2\rangle_E \\ U_{E2}U_{AB}^{5}U_{E1}(q_A^i, q_B^i, e^i) = |\Phi^+\rangle_{AB}|f_1\rangle_E + |\Psi^+\rangle_{AB}|f_3\rangle_E \\ U_{E2}U_{AB}^{6}U_{E1}(q_A^i, q_B^i, e^i) = -|\Phi^-\rangle_{AB}|f_2\rangle_E + |\Psi^-\rangle_{AB}|f_4\rangle_E \\ U_{E2}U_{AB}^{7}U_{E1}(q_A^i, q_B^i, e^i) = -|\Phi^-\rangle_{AB}|f_2\rangle_E - |\Psi^-\rangle_{AB}|f_4\rangle_E \\ U_{E2}U_{AB}^{8}U_{E1}(q_A^i, q_B^i, e^i) = |\Phi^+\rangle_{AB}|f_1\rangle_E - |\Psi^+\rangle_{AB}|f_3\rangle_E \end{cases} \quad (41)$$

The equation group (41) is the result of the collective attack on the proposed LAQKD protocol 3. TP tries to use the measurement result of his/her probe quantum system $E$ in equation group (9) to obtain the final shared key $K$ and the pre-shared key $K_1$. For easily understanding of the collective attack result, we re-express the equation group (41) and Table 3 into Table 5. Similar to the proof of Theorem 1, according to Table 5, we can find that no matter which measurement result of $E$ is obtained by TP, he/she can obtain nothing of $R_A$ and $K_1$. The Theorem 3 is proved.

Table 5. Collective attack on the proposed LAQKD protocol 3

| MR of $E$ | MR of $\{Q_A', Q_B'\}$ | $H_A', H_B'$ | $R_A$ | $K_1$ |
|---|---|---|---|---|
| $|f_1\rangle_E$ | $|\Phi^+\rangle_{AB}$ | $H'^0, H'^0$ | 0 | 0 |

|   |   | $H'^2, H'^2$ | 1 | 0 |
|---|---|---|---|---|
|   |   | $H'^1, H'^1$ | 0 | 1 |
|   |   | $H'^3, H'^3$ | 1 | 1 |
| $\|f_2\rangle_E$ | $\|\Phi^-\rangle_{AB}$ | $H'^0, H'^0$ | 0 | 0 |
|   |   | $H'^2, H'^2$ | 1 | 0 |
|   |   | $H'^1, H'^3$ | 0 | 1 |
|   |   | $H'^3, H'^1$ | 1 | 1 |
| $\|f_3\rangle_E$ | $\|\Psi^+\rangle_{AB}$ | $H'^0, H'^2$ | 0 | 0 |
|   |   | $H'^2, H'^0$ | 1 | 0 |
|   |   | $H'^1, H'^1$ | 0 | 1 |
|   |   | $H'^3, H'^3$ | 1 | 1 |
| $\|f_4\rangle_E$ | $\|\Psi^-\rangle_{AB}$ | $H'^0, H'^2$ | 0 | 0 |
|   |   | $H'^2, H'^0$ | 1 | 0 |
|   |   | $H'^1, H'^3$ | 0 | 1 |
|   |   | $H'^3, H'^1$ | 1 | 1 |

The above three theorems and the corresponding proofs show that all the proposed three LAQKD protocols are robust.

## 4. Key recycling

Section 3 shows that, in the three proposed LAQKD protocols, no eavesdropper can obtain the pre-shared keys $K_1$ and $K_2$ without being detected. That means, if no eavesdropper has been detected in the protocol, the pre-shared keys can be reused next time. However, if the participants find an eavesdropper during running the protocol, do the pre-shared keys still can be reused? In most of the existing authenticated quantum protocols [17, 24-26], if an eavesdropper has been detected, all the pre-shared keys must be discarded. Different from these existing authenticated quantum protocols, however, the proposed LAQKD protocols can recycle parts of the pre-shared keys when an eavesdropper has been detected. This section gives the expectation of the key recycling rate. Here, for a pre-shared key $K_1$, the key recycling rate $rate(K_1)$ is defined as follows.

$$rate(K_1) = \frac{bit(K_1) - leakage(K_1)}{bit(K_1)}$$

The $bit(K_1)$ is the number of bits of the whole $K_1$ and the $leakage(K_1)$ is the maximal leakage bits in the protocol.

According to the key recycling rate, if an eavesdropper has been detected in the protocol, the involved participants just need to discard parts of the pre-shared keys and the remaining parts still can be used in next time.

Before analyzing the key recycling rate in each proposed LAQKD protocol, some background is introduced here.

**4.1 Background of key recycling**

The [27] pointed out that there is a maximal probability of 85.4% to obtain the correct value of the four single photons $\{|0\rangle, |1\rangle, |+\rangle, |-\rangle\}$. That is, suppose that the values of single photons $\{|0\rangle, |+\rangle\}$ are equal to '0' and the values of single photons $\{|1\rangle, |-\rangle\}$ are equal to '1'. Then, for a single photon sequence $S = \{s_1, s_2, \cdots, s_n\}$ where each photon $s_i$ is randomly chosen from $\{|0\rangle, |1\rangle, |+\rangle, |-\rangle\}$, the maximal probability of obtaining the correct value of $s_i$ is 85.4%. According to Shannon's entropy $H(X) = -\sum_{x} p(x)\log_2 p(x)$ where $p(x)$ is the probability of possible values $x$, the maximal information leakage of the value of $s_i$ is 0.41 bit. Besides, the $s_i$ will collapse after obtaining the value of it.

The [28] showed that, without knowing the exact value of $s_i$, the basis of $s_i$ cannot be obtained. Moreover, if the value of $s_i$ is known, the maximal information leakage of $s_i$'s basis is 0.40 bit. For example, assume there is a classical bit $a \in \{0,1\}$ and $s_i$ is generated according to $a$. That is, if $a=0$, then $s_i=|0\rangle$. Otherwise, $s_i=|+\rangle$. Theoretically, the maximal information entropy of $a$ obtained from $s_i$ is 0.4 bit. Similarly, the $s_i$ will collapse after obtaining the basis of it.

According to the above two studies [27] [28], we can get the following lemma.

**Lemma 1:** For one qubit which is randomly chosen from the four single states $\{|0\rangle, |1\rangle, |+\rangle, |-\rangle\}$, the basis of this qubit cannot be obtained from this qubit itself.

**Proof:** Assume $s_i$ is a qubit randomly chosen from $\{|0\rangle, |1\rangle, |+\rangle, |-\rangle\}$. If we try to obtain $s_i$'s basis from the qubit itself, according to the study [28], the following two conditions must be met simultaneously.

(1) The exact value of $s_i$ must have been obtained.

(2) The quantum particle $s_i$ has not collapsed yet.

However, according to the study [27], we can find that if we obtain $s_i$'s value from the qubit itself, then this qubit will collapse. Hence, the above two conditions cannot be satisfied at the same time. As a result, we cannot obtain $s_i$'s basis from the qubit itself. The Lemma 1 is proved.

## 4.2 Key recycling rate of $K_1$ in the proposed LAQKD protocol 1

According to Lemma 1, in the proposed LAQKD protocol 1, if an eavesdropper TP wants to obtain the pre-shared key $K_1$, he/she must obtain the exact values of $\{q_A^{ri}, q_B^{ri}\}$ first. In $\{q_A^{ri}, q_B^{ri}\}(1 \leq i \leq n+m)$, the values of $\{q_A^{ri}, q_B^{ri}\}(1 \leq i \leq n)$ are decided by $\{R_A, R_B\}$ and the values of $\{q_A^{ri}, q_B^{ri}\}(n+1 \leq i \leq n+m)$ are decided by $\{h(R_A), h(R_B)\}$ where $\{h(R_A), h(R_B)\}$ are derived from $\{R_A, R_B\}$. Here, when $\{R_A, R_B\}$ and $\{h(R_A), h(R_B)\}$ are considered separately, both of them can be considered as random. Hence, TP cannot obtain the bases of $\{q_A^{ri}, q_B^{ri}\}(1 \leq i \leq n)$ or the bases of $\{q_A^{ri}, q_B^{ri}\}(n+1 \leq i \leq n+m)$ separately. According to this, TP just has two strategies to obtain $K_1$ as follows.

Strategy 1: TP obtains parts information of $\{R_A, R_B\}$ from $\{q_A^{ri}, q_B^{ri}\}(1 \leq i \leq n)$ and then uses the obtained information to deduce parts of $K_1$ from $\{q_A^{ri}, q_B^{ri}\}(n+1 \leq i \leq n+m)$.

Strategy 2: TP obtains parts information of $\{h(R_A), h(R_B)\}$ from $\{q_A^{ri}, q_B^{ri}\}(n+1 \leq i \leq n+m)$ and then uses the obtained information to obtain parts of $K_1$ from $\{q_A^{ri}, q_B^{ri}\}(1 \leq i \leq n)$.

In Strategy 1, the maximal information leakage of $\{R_A, R_B\}$ from $\{q_A^{ri}, q_B^{ri}\}(1 \leq i \leq n)$ is $2 \times 0.41n = 0.82n$ bits. The expected information leakage of $\{h(R_A), h(R_B)\}$ from the obtained $0.82 \times n$ bits is $0.82 \times n \times \frac{m}{n} = 0.82m$ bits. Hence, we can consider that TP obtains $0.82m$-bit values of $\{q_A^{ri}, q_B^{ri}\} (n+1 \leq i \leq n+m)$. Then, for per leakage bit, TP maximally can obtain 0.4 bit of $K_1$. Hence, the maximal information leakage of $K_1$ from the whole $\{q_A^{ri}, q_B^{ri}\}(1 \leq i \leq n+m)$ is $0.4 \times 0.82m \approx 0.33m$ bits.

In Strategy 2, the maximal information leakage of $\{h(R_A), h(R_B)\}$ from $\{q_A^{ri}, q_B^{ri}\}(n+1 \leq i \leq n+m)$ is $0.82m$ bits. All the leakage bits can be considered as the value leakage of $\{q_A^{ri}, q_B^{ri}\}(1 \leq i \leq n)$. Then, the maximal information leakage of $K_1$ still is $0.33m$.

Hence, no matter which attack strategy is chosen by TP, the maximal information leakage of $K_1$ is $0.33m$ bits. The key recycling rate $rate(K_1)$ is

$$rate(K_1) = \frac{bit(K_1) - leakage(K_1)}{bit(K_1)} = \frac{n+m-0.33m}{n+m} = \frac{n+0.67m}{n+m}.$$

**4.3 Key recycling rate of $K_1$ and $K_2$ in the proposed LAQKD protocol 2**

Different from Protocol 1, in Protocol 2, the $\{h(R_A), h(R_B)\}$ are directly announced. Hence, the information leakage of $\{R_A, R_B\}$ can be considered as $2m$ bits and then the information leakage of $K_1$ is $0.4 \times 2m = 0.8m$ bits. Then, the key recycling rate of $K_1$ is $rate(K_1) = \frac{bit(K_1) - leakage(K_1)}{bit(K_1)} = \frac{n-0.8m}{n}$.

For the $K_2$ which is used to choose the positions for dividing $R$ into $R_A$ and $R_B$. TP just can obtain the information of $K_2$ from the announced $\{h(R_A), h(R_B)\}$. Hence, the maximal information leakage of $K_2$ is $2m$ bits. The key recycling rate of $K_2$ is $rate(K_2) = \frac{bit(K_2) - leakage(K_2)}{bit(K_2)} = \frac{n-2m}{n}$.

**4.4 Key recycling rate of $K_1$ in the proposed LAQKD protocol 3**

The key recycling rate of $K_1$ in protocol 3 is the same as it in protocol 1. That is, because $K_1$ in protocol 3 is used to perform $H'$ on each qubit, $K_1$ can be considered as the bases of $\{q_A^{ri}, q_B^{ri}\}(1 \leq i \leq n+m)$. Moreover, the values of $\{q_A^{ri}, q_B^{ri}\}(1 \leq i \leq n+m)$ are decided by the $\{R_A \| h(R_A), R_B \| h(R_B)\}$. Here, the decision methods of both the value and bases of $\{q_A^{ri}, q_B^{ri}\}(1 \leq i \leq n+m)$ are the same as those in the proposed Protocol 1. Hence, the key recycling rate of $K_1$ is the same as it in the proposed Protocol 1 where $rate(K_1) = \frac{n+0.67m}{n+m}$.

# 5. Comparison

This section compares the three proposed LAQKD protocols with several existing authenticated quantum key distribution protocols [16, 22, 26]. Before comparing, a new concept named transmission time cost should be introduced first here.

**5.1 Transmission time cost of a quantum protocol**

For a quantum protocol, the transmission time cost is used to quantify the qubits and the classical bits transmission time needed for running the protocol. According to our knowledge, this study is the first one to propose the transmission time cost concept for quantum protocols. Most of the existing quantum protocols just focus on getting a better qubit efficiency, needing fewer quantum capabilities or using some quantum resources which are easier for implementation. However, to achieve these requirements, several existing quantum protocols need to cost a lot of the transmission time.

For example, Sun et al.'s [29] proposed an improvement on Liu et al.'s [30]

quantum key agreement protocol. In Liu et al.'s protocol, every involved participant needs to generate and sends a qubit sequence to every other participant. In Sun et al.'s improved protocol, every participant just needs to generate one qubit sequence and then transmits it among all the involved participants one by one. Though Sun et al.'s improvement has a higher qubit efficiency by its qubit transmission method, this qubit transmission method must cost a lot of time. That is, in Liu et al.'s protocol, all the qubit transmission processes can be carried out simultaneously. Hence, if assuming that the time cost of waiting for qubit transmission one time is '1', then the time cost of qubit transmission processes in Liu et al.'s protocol is only '1'. In Sun et al.'s protocol, for one qubit sequence, it should be transmitted $N-1$ times where $N$ is the number of involved participants. According to this, the qubit transmission time cost of Sun et al.'s protocol is '$N-1$'. Hence, though the higher qubit efficiency Sun et al.'s improved protocol has, Liu et al.'s protocol has the less qubit transmission time cost. So it is difficult to determine which protocol is more efficient. To make the comparison of quantum protocols more comprehensive, the transmission time cost should be considered.

The transmission time cost ($TTC$) of a protocol is defined as 'the minimum waiting time of the qubit transmission and classical bit transmission for running a protocol'. Here, the 'minimum' means that if several transmission processes can be carried out simultaneously, then these processes are just considered as one time. Assume that the time cost of transmitted qubits is the same as the time cost of transmitted classical bits, then $TTC = TTC(q) + TTC(b) - TTC(q\&b)$ where $TTC(q)$ and $TTC(b)$ are the minimum waiting time of all the qubit transmission processes and the classical bit transmission processes, respectively. $TTC(q\&b)$ denotes some transmission processes where both the qubits and the classical bits can be transmitted simultaneously.

**5.2 Comparison of the proposed protocols and several existing protocols**

In this part, the comparisons of the three proposed LAQKD protocols and several existing quantum key distribution protocols are shown. Because this study is the first to satisfy both the authentication and all participants are lightweight. Hence, there are no similar protocols to compare with the proposed protocols. To solve this problem, the comparison is divided into two parts. First, in Table 6, the proposed protocols are compared with a lightweight quantum key distribution (LQKD) protocol [16]. Second, in Table 7, the comparison among the proposed protocols and two authenticated quantum key distribution (AQKD) protocols is given [22, 26].

The comparison includes the qubit efficiency, the bit number of pre-shared master keys, the quantum resources, the quantum capabilities, the key recycling and the transmission time cost.

**Qubit efficiency (QE)**: The qubit efficiency is used to quantify the number of quantum particles needed for sharing a secure session key. Several existing quantum protocols [22, 26] use the equation $QE = \frac{b}{q}$ to quantify it. Here, $QE$ is the abbreviation of the qubit efficiency, $b$ denotes the bits of final shared raw key and $q$ is the total number of used quantum particles in the protocol. The less $QE$ the better qubit efficiency is.

**Bit number of pre-shared master keys (PSK)**: The bit number of pre-shared master keys is used to quantify the cost of key management for running a protocol. The pre-shared master keys in each protocol mean that the involved participants need to hold the keys for a long time for running a protocol when they need.

**Quantum resources (QR):** Quantum resources is used to denote what quantum states are needed for running a protocol. Such as single photon, Bell state, Cluster state and so on.

**Quantum capabilities (QC):** Quantum capabilities are used to describe the quantum capabilities of each participant and TP needed for running a protocol.

**Key recycling (KR):** The key recycling rate is used to quantify how many bits of the pre-shared master keys can be recycled when an eavesdropper has been detected. Here, if a protocol discards all the pre-shared master keys when an eavesdropper has been detected, then the key recycling rate is denoted as N/A.

**Transmission time cost (TTC):** It is used to quantify the time needed for running a protocol as shown above.

Table 6. Comparison among LQKD protocols

|  | Hwang et al.[16] | Proposed protocol 1 | Proposed protocol 2 | Proposed protocol 3 |
|---|---|---|---|---|
| **QE** | $\frac{1}{9}$ | $\frac{1}{2} \times \frac{n}{n+m}$ | $\frac{1}{2}$ | $\frac{1}{2} \times \frac{n}{n+m}$ |
| **QR** | Bell state | Single photon | Bell state | Single photon |
| **QC of TP** | 1. Bell measurement  2. Generate Bell state | 1. Bell measurement | 1. Bell measurement  2. Generate Bell state | 1. Bell measurement  2. Generate single photon |
| **QC of participants** | 1. Unitary operations  2. Reflect | 1. Unitary operations  2. Generate | 1. Unitary operations  2. Measure | 1. Unitary operations  2. Reflect |
| **TTC** | 5 | 2 | 2 | 3 |

Table 7. Comparison among AQKD protocols

|  | Li et al.[26] | Tsai et al.[22] | Proposed protocol 1 | Proposed protocol 2 | Proposed protocol 3 |
|---|---|---|---|---|---|
| **QE** | $\frac{2}{9} \times \frac{n}{n+m}$ | $\frac{1}{4}$ | $\frac{1}{2} \times \frac{n}{n+m}$ | $\frac{1}{2}$ | $\frac{1}{2} \times \frac{n}{n+m}$ |
| **PSK** | $3(n+m)$ | $2n$ | $n+m+l$ | $2(n+m+l)$ | $n+m+l$ |

| QR | Bell state | Single photon | Single photon | Bell state | Single photon |
| --- | --- | --- | --- | --- | --- |
| KR | N/A | N/A | $\frac{n+0.67m}{n+m}$ | $K_1: \frac{n-0.8m}{n}$ $K_2: \frac{n-2m}{n}$ | $\frac{n+0.67m}{n+m}$ |
| TTC | 2 | 3 | 2 | 2 | 3 |

## 6. Conclusions

To make the quantum key distribution protocols more practical, this paper designs three different LAQKD protocols from two perspectives: lightweight and authenticated. Moreover, in different proposed protocols, the different lightweight quantum capabilities of the involved participants are required. Hence, with these proposed protocols, the participants can flexibly choose a suitable protocol to use according to their own lightweight quantum capabilities. Besides, different from most of the existing authenticated quantum protocols, a key recycling threshold is given for each proposed LAQKD protocol. When the eavesdropping is detected, the participants need not to discard all the pre-shared keys and share pre-shared keys again, they can recycle most parts of the pre-shared keys instead. This point also makes the proposed protocols more practical than others. Furthermore, a new concept named transmission time cost is proposed in this paper. This concept makes the comparison of quantum protocols more comprehensive.

By the way, considering the real environment, all the protocols proposed in this paper have a disadvantage. That is, all the participants involved in these protocols are required to have the same lightweight quantum capabilities. For example, in the proposed LAQKD protocol 1, both Alice and Bob need to have the same quantum capabilities: generating Z-basis qubits and performing single photon unitary operations. Therefore, how to help two participants with different lightweight quantum capabilities share keys is still a problem worthy of research.

# Acknowledgment

We would like to thank the Ministry of Science and Technology of the Republic of China, Taiwan for partially supporting this research in finance under the Contract No. MOST 109-2221-E-006-168-; No. MOST 108-2221-E-006-107-.